\documentclass[conference]{IEEEtran}
\IEEEoverridecommandlockouts
\usepackage{cite}
\usepackage{amsmath,amssymb,amsfonts,bm}
\usepackage{graphicx}
\usepackage[nolist]{acronym}
\usepackage{textcomp}
\usepackage[bottom]{footmisc}
\usepackage[table,dvipsnames]{xcolor}
\usepackage{tikz}%Draw schematics
\usepackage{hyperref}
\usetikzlibrary{positioning,shapes,matrix,fit,backgrounds,arrows,chains,patterns}
\usepackage{pgfplots} % Graphs
\usepgfplotslibrary{fillbetween} %Fill below graphs
\usepackage[outline]{contour}
\usepackage[commentColor=black]{algpseudocodex}
\usepackage{multirow}
\tikzset{algpxIndentLine/.style={draw=black}}
\algrenewcommand{\alglinenumber}[1]{\bfseries\footnotesize #1}
\algrenewcommand{\textproc}{}
\algrenewcommand{\algorithmicrequire}{\textbf{Input:}}
\algrenewcommand{\algorithmicensure}{\textbf{Output:}}
\usepackage{algorithm}
\usepackage{caption} 
\usepackage{subcaption} %side-by-side figures with sub-captions

\def\BibTeX{{\rm B\kern-.05em{\sc i\kern-.025em b}\kern-.08em
    T\kern-.1667em\lower.7ex\hbox{E}\kern-.125emX}}

\newcommand\copyrighttext{%
    \footnotesize \textcopyright 2025 IEEE. Personal use of this material is permitted. Permission from IEEE must be obtained for all other uses, in any current or future media, including reprinting/republishing this material for advertising or promotional purposes, creating new collective works, for resale or redistribution to servers or lists, or reuse of any copyrighted component of this work in other works. DOI: \href{https://ieeexplore.ieee.org/document/11184612}{10.1109/TCOMM.2025.3615681}
}
\newcommand\copyrightnotice{%
\begin{tikzpicture}[remember picture,overlay]
\node[anchor=south,yshift=10pt] at (current page.south) {\fbox{\parbox{\dimexpr\textwidth-\fboxsep-\fboxrule\relax}{\copyrighttext}}};
\end{tikzpicture}%
}

\begin{document}

\begin{acronym}
  \acro{ML}{Machine Learning}
  \acro{AI}{Artificial Intelligence}
  \acro{NN}{Neural Network}
  \acro{DNN}{Deep Neural Network}
  \acro{BP}{Belief Propagation}
  \acro{NBP}{Neural Belief Propagation}
  \acro{GNBP}{Gated Neural Belief Propagation}
  \acro{ECC}{Error Correcting Codes}
  \acro{RL}{Reinforcement Learning}
  \acro{LDPC}{Low Density Parity Check}
  \acro{DL}{Deep Learning}
  \acro{BLER}{Block Error Rate}
  \acro{SGD}{Stochastic Gradient Descent}
  \acro{SNR}{Signal to Noise Ratio}
  \acro{AWGN}{Additive White Gaussian Noise}
  \acro{LLR}{Log Likelihood Ratio} 
  \acro{RNN}{Recurrent Neural Network}
  \acro{BCE}{Binary Cross-Entropy}
  \acro{SGD}{Stochastic Gradient Descent}
  \acro{CDF}{Cumulative Density Function}
  \acro{PEG}{Progressive Edge Growth}
  \acro{DSF}{Differentiable Step Function}
  \acro{BSC}{Binary Symmetric Channel}
  \acro{DLL}{Discrete Levels Learning}
  \acro{PC}{Parity-Check}
\end{acronym}

\title{Learning Linear Block Codes with Gradient Quantization}

\author{
    \IEEEauthorblockN{Louis-Adrien Dufrène\textsuperscript{\textsection}, Quentin Lampin\textsuperscript{\textsection}, Guillaume Larue\textsuperscript{\textsection}}
    \IEEEauthorblockA{Orange Research, Meylan, France\\
    Email: \{louisadrien.dufrene, quentin.lampin, guillaume.larue\}@orange.com}
}

\maketitle
\copyrightnotice
\begingroup\renewcommand\thefootnote{\textsection}
\footnotetext{The order of author names does not reflect the level of contribution; all authors contributed equally to this paper.}
\endgroup

\begin{abstract}
This study investigates the problem of learning linear block codes optimized for Belief-Propagation decoders significantly improving performance compared to the state-of-the-art. Our previous research is extended with an enhanced system design that facilitates a more effective learning process for the parity check matrix. We simplify the input dataset, restrict the number of parameters to learn and improve the gradient back-propagation within the model. We also introduce novel optimizers specifically designed for discrete-valued weights. Based on conventional gradient computation, these optimizers provide discrete weights updates, enabling finer control and improving explainability of the learning process. Through these changes, we consistently achieve improved code performance, provided appropriately chosen hyper-parameters. To rigorously evaluate the performance of learned codes in the context of short to medium block lengths, we propose a comprehensive code performance assessment framework. This framework enables a fair comparison between our learning methodology and random search approaches, ensuring statistical significance in our results. The proposed model pave the way for a new approach to the efficient learning of linear block codes tailored to specific decoder structures.
\end{abstract}


\section{Introduction}
\label{section:intro}
The automatic learning of \ac{ECC} is an open research subject, rising substantial interest within the research community. While the construction of efficient codes is intricately linked with expert information theory knowledge and practical engineering considerations, the advent of \ac{ML} enable innovative code design strategies. The expectations range from the discovery of new construction methods for existing code families to the design of novel code structures with desirable properties.  

For instance, in \cite{KO-Codes}, the authors employ \ac{DL} techniques to design a novel code family known as KO (Kronecker Operation) codes. Drawing inspiration from the Kronecker operations used as the foundation of Reed-Muller and Polar codes, they replace the conventional Plotkin trees with a trainable \ac{NN}. After an end-to-end training, this approach results in novel non-linear encoding and decoding functions. In a similar manner, \cite{TurboAE} uses an end-to-end \ac{DL} method to learn new efficient codes, based on the Turbo codes encoder and decoder architectures. These novel non-linear structured codes demonstrate the capability of \ac{ML} to search for new codes design.

More conventionally, \ac{ML} is used to improve the performance of codes belonging to existing codes families. In \cite{RL-Codes}, three different \ac{RL} techniques are used to learn codes: a policy gradient, a genetic algorithm and an Advantage Actor Critic (A2C). These learned codes are structured using known codes families, e.g. linear block codes based on a binary generator matrix, and the study proposes a generic constructor-evaluator approach to learn them with \ac{RL} methods. In the same vein, \cite{RL-LDPC} proposes a genetic algorithm to learn the parity check matrix of \ac{LDPC} codes. A decoder is used to evaluate the reward provided to the \ac{RL} agent. In the short block length regime, the genetic algorithm delivers well-designed codes tailored to the specific decoder in use, outperforming traditional design tools that often fail to produce efficient codes in this context. Some unconventional choices for \ac{LDPC} design, such as degree-1 variable nodes, also demonstrate the potential of \ac{ML} to expand our insights into code construction.

In addition to \ac{RL} techniques, both supervised and unsupervised \ac{DL} are also employed to enhance the performance of known code structures. In \cite{DL-Polar}, a gradient descent based optimization is used to learn the position of the information/frozen bits indices of Polar codes within a \ac{BP} decoder. The trainable weights are solely the bits position. The resulting codes have competitive performances when compared to the 5G Polar code with the same decoder. In \cite{Choukroun2024a}, a unified encoder-decoder training framework is proposed to learn binary linear block codes, where both the code and a Transformer-based decoder are jointly optimized in an end-to-end fashion. In \cite{Choukroun2024b} the same authors apply back-propagation methods over a differentiable Tanner Graph to optimize codes specifically for BP decoding. In a previous work \cite{GNBP_Learning}, we proposed an auto-encoder to efficiently learn linear block codes with a \ac{BP}-based decoding. The auto-encoder is composed of a differentiable encoder and decoder, used to learn the parity-check matrix. The decoder is a \ac{GNBP}, a less-complex variant of the \ac{NBP} proposed in \cite{Nachmani_NBP}, with an iterative graph decoding structure similar to a weighted \ac{BP}.

In this paper, we propose to continue the work started in \cite{GNBP_Learning}. Our work specifically targets applications where limited computational resources and energy constraints, or real-time operations, necessitate efficient decoding with few iterations. This is particularly relevant for Internet of Things (IoT) and Ultra Reliable Low Latency Communications (URLLC) systems, where short packets and efficient decoding are essential requirements. BP decoding, despite its sub-optimality for short dense codes, remains highly attractive for practical implementations due to its widespread, fully parallelizable and scalable architecture, with efficient hardware support. Moreover, long \ac{LDPC} codes with \ac{BP} decoding - as seen in 5G data channels - already achieve excellent performance-complexity trade-offs due to their sparse structure. This also motivates our focus on learning short-to-medium length codes specifically optimized for BP decoding, aiming to enable a unified decoding architecture, performing well across all length regimes. Our goal is then not to find theoretically optimal codes but rather short codes that perform well under BP decoding. A novel architecture is described, simplifying the overall \ac{NN} model and learning process, focusing solely on learning the parity check matrix. When the appropriate hyper-parameters are selected, we demonstrate a rapid and consistent learning, that results in codes showcasing superior performance.

The paper is organized as follow: In Section~\ref{section:system_model}, we present the system model, outlining the novel auto-encoder architecture used for learning the codes. Section~\ref{section:code_comparison_methodology} presents a new framework to validate the efficiency of learning methodologies by verifying that learned codes significantly outperform random ones. Section~\ref{section:gradient_quantificatin_mechanisms} introduces a gradient quantization mechanism and the associated discrete optimizer, which provide a novel approach to handle the binary-valued weights in the learning process. In Section~\ref{section:training_parameters}, we detail the training hyper-parameters and methodologies employed in our experiments. The simulation results are presented in Section~\ref{section:results}, showcasing the performance of the learned codes. Finally, Section~\ref{section:conclusion} concludes the paper and discuss potential directions for future works.

\section{Auto-Encoder Model \& Training}
\label{section:system_model}
\begin{figure}%[htbp]
    \centerline{\usetikzlibrary{fit,backgrounds}
\contourlength{1.4pt}
\tikzset{>=latex} % for LaTeX arrow head

\begin{tikzpicture}[scale=0.725,every node/.style={transform shape} ] %every node/.style={scale=0.9}]
    %Manually reduce line spacing to single space no matter double or single column document (draft mode)
    \linespread{1}
    %
    
    % Input
    \node (in) at (0,0) {};
    \node [above= 0.05em of in] {$(\overbrace{\begin{matrix}
        1 & \hdots & 1\\
    \end{matrix}}^{N_\mathrm{valid}}|\overbrace{\begin{matrix}
        -1 & \hdots & -1\\
    \end{matrix}}^{N_\mathrm{errors}})$};

    % Error Vector
    %\node [draw,line width=.8pt,rounded corners, below= 3em of in, minimum width=4cm,minimum height=0.6cm] (EV) {Error Vector};
    %\draw[very thick,-latex,rounded corners] (in) -- (EV);

    % Shuffle
    \path (in) ++ (+0,-1.5) coordinate (Sh0);
    \path (Sh0) ++ (+0.25,-0.25) coordinate (Sh1);
    \path (Sh1) ++ (-0,-0.3) coordinate (Sh2);
    \path (Sh2) ++ (-0.5,-0.3) coordinate (Sh3);
    \path (Sh3) ++ (-0,-0.5) coordinate (Sh4);

    \path (Sh0) ++ (-0.25,-0.25) coordinate (Sh5);
    \path (Sh5) ++ (-0,-0.3) coordinate (Sh6);
    \path (Sh6) ++ (+0.5,-0.3) coordinate (Sh7);
    \path (Sh7) ++ (-0,-0.5) coordinate (Sh8);

    %\path (Sh0) ++ (0,+0.) coordinate (ShuffleC);
        
    \draw[very thick,-latex,rounded corners, dashed] (Sh1) -- (Sh2)-- (Sh3) -- (Sh4);
    \draw[very thick,-latex,rounded corners] (Sh5) -- (Sh6)-- (Sh7) -- (Sh8);

    \node [] (Shuffle) at (Sh0) {Random Shuffle};
    
    %\node [draw,line width=.8pt,rounded corners, below= 1.5em of EV, minimum width=4cm,minimum height=0.6cm] (Shuffle) {Shuffle};

    \node [] (Sh1) at (Sh1) {};
    \node [] (Sh8) at (Sh8) {};
    \node (BoxShuffle) [draw,line width=.8pt, rounded corners,fit = (Shuffle) (Sh1) (Sh8), scale=1] {};

    \draw[very thick,-latex,rounded corners] (in) to node[right] {    Shape: $[-1,n]$} (BoxShuffle);
    %AWGN
    %\node [thick,circle,draw, inner sep=0.5,outer sep=0.6, below= 1.5em of Shuffle] (AWGN) {\Large $+$};
    %\node [align=center, right= 2em of AWGN] (NoiseLvl) {$\mathcal{N}(0,N_0/2)$};
    %\node [align=center, left= 0.5em of AWGN] (AWGNLabel) {AWGN};
    %\draw[very thick,-latex,rounded corners] (Shuffle) -- (AWGN);
    %\draw[very thick,-latex,rounded corners] (NoiseLvl) -- (AWGN);
    
    % sCALING
    \node [draw,line width=.8pt,rounded corners,below= 1.5em of BoxShuffle, minimum width=4cm,minimum height=0.6cm] (Scaling) {Scaling $\times \alpha$};
    \draw[very thick,-latex,rounded corners] (BoxShuffle) -- (Scaling);
    
    % Broadcast
    \path (Scaling.south) ++ (0,-1) coordinate (ScalingS1);
    \node [draw,line width=.8pt,rounded corners,minimum width=7cm,minimum height=0.6cm] (BC) at (ScalingS1) {Iteration Broadcast};
    \path (BC.south) ++ (+3.5,0) coordinate (BCS0);
    \path (BC.south) ++ (+2.5,0) coordinate (BCS1);
    \path (BC.south) ++ (+1.5,0) coordinate (BCS2);
    \path (BC.south) ++ (+0.5,0) coordinate (BCS3);
    \path (BC.south) ++ (-0.5,0) coordinate (BCS4);
    \path (BC.south) ++ (-1.5,0) coordinate (BCS5);
    \path (BC.south) ++ (-2.5,0) coordinate (BCS6);
    \path (BC.south) ++ (-3.5,0) coordinate (BCS7);
    %\path (BC.south) ++ (-3.5,0) coordinate (BCS75);
    %\path (BC.south) ++ (-4,0) coordinate (BCS8);
    
    \path (BC.north) ++ (0,0) coordinate (BCN3);
    \draw[very thick,-latex,rounded corners] (Scaling) to node[right] {${\bm{\lambda}}$} (BCN3);
    
    \node [below right=1em of BCS5] (BCDots) {$\hdots$};
    
    % Iteration Weighting
    \path (BC.south) ++ (+0,-1) coordinate (BCS);
    %\node [draw,line width=.8pt,rounded corners,minimum width=7cm,minimum height=0.6cm, fill=gray!50] (IW) at (BCS) {Iteration Normalization \& Weighting};
    %\path (IW.north) ++ (+3.5,0) coordinate (IWN0);
    %\path (IW.north) ++ (+2.5,0) coordinate (IWN1);
    %\path (IW.north) ++ (+1.5,0) coordinate (IWN2);
    %\path (IW.north) ++ (+0.5,0) coordinate (IWN3);
    %\path (IW.north) ++ (-0.5,0) coordinate (IWN4);
    %\path (IW.north) ++ (-1.5,0) coordinate (IWN5);
    %\path (IW.north) ++ (-2.5,0) coordinate (IWN6);
    %\path (IW.north) ++ (-3.5,0) coordinate (IWN7);
    
    %\path (IW.south) ++ (+3.5,0) coordinate (IWS0);
    %\path (IW.south) ++ (+2.5,0) coordinate (IWS1);
    %\path (IW.south) ++ (+1.5,0) coordinate (IWS2);
    %\path (IW.south) ++ (+0.5,0) coordinate (IWS3);
    %\path (IW.south) ++ (-0.5,0) coordinate (IWS4);
    %\path (IW.south) ++ (-1.5,0) coordinate (IWS5);
    %\path (IW.south) ++ (-2.5,0) coordinate (IWS6);
    %\path (IW.south) ++ (-3.5,0) coordinate (IWS7);
    
    %\draw[very thick,-latex,rounded corners] (BCS1) to node[right] {$\bm{\lambda}$} (IWN1);
    %\draw[very thick,-latex,rounded corners] (BCS3) to node[right] {$\bm{\lambda}$} (IWN3);
    %\draw[very thick,-latex,rounded corners] (BCS6) to node[right] {$\bm{\lambda}$} (IWN6);
    
    %\node [below right=1em of IWS5] (IWDots) {$\hdots$};
    
    % RNN Decoder
    \node [draw,line width=.8pt,rounded corners, text width=1cm,minimum height=0.6cm, align=center, below=2.25em of BCS1] (GNBP0) {BP RNN Cell};
    \node [draw,line width=.8pt,rounded corners, text width=1cm,minimum height=0.6cm, align=center, below=2.25em of BCS3] (GNBP1) {BP RNN Cell};
    \node [draw,line width=.8pt,rounded corners, text width=1cm,minimum height=0.6cm, align=center, below=2.25em of BCS6] (GNBP5) {BP RNN Cell};
    
    \draw[very thick,-latex,rounded corners] (BCS1) to node[right] {$\bm{\lambda}$} (GNBP0);
    \draw[very thick,-latex,rounded corners] (BCS3) to node[right] {$\bm{\lambda}$} (GNBP1);
    \draw[very thick,-latex,rounded corners] (BCS6) to node[right] {$\bm{\lambda}$} (GNBP5);
    
    \draw[very thick,-latex,rounded corners] (GNBP0) -- (GNBP1);
    \draw[very thick,-latex,rounded corners,dashed] (GNBP1) to node[above] {$\mathbf{\mu}_{c \xrightarrow[]{} v}$} (GNBP5);
    
    \path (GNBP5.west) ++ (-0.25,+0) coordinate (RNNLabelC);
    \path (GNBP0.east) ++ (+0.25,+0) coordinate (RNNLabelCPhantom);
    \node [align=center, rotate=90] (RNNLabel) at (RNNLabelC) {\textbf{RNN}};
    \node [align=center, rotate=90] (RNNLabelPhantom) at (RNNLabelCPhantom) {\textbf{}};
    \node (BoxRNN) [draw,line width=.8pt,dashed, rounded corners,fit = (GNBP0) (GNBP5) (RNNLabel) (RNNLabelPhantom), scale=1] {};

    % Iteration output Weighting
    %\path (GNBP1.south) ++ (0,-1.25) coordinate (GNBP1S);
    \path (BC.south) ++ (+0,-3.5) coordinate (BCS);
    \node [draw,line width=.8pt,rounded corners,minimum width=7cm,minimum height=0.6cm] (IOW) at (BCS) {Iteration Output Recombination};
    
    \path (IOW.north) ++ (+3.5,0) coordinate (IOWN0);
    \path (IOW.north) ++ (+2.5,0) coordinate (IOWN1);
    \path (IOW.north) ++ (+1.5,0) coordinate (IOWN2);
    \path (IOW.north) ++ (+0.5,0) coordinate (IOWN3);
    \path (IOW.north) ++ (-0.5,0) coordinate (IOWN4);
    \path (IOW.north) ++ (-1.5,0) coordinate (IOWN5);
    \path (IOW.north) ++ (-2.5,0) coordinate (IOWN6);
    \path (IOW.north) ++ (-3.5,0) coordinate (IOWN7);
    %\path (IOW.north) ++ (-3.5,0) coordinate (IOWN75);
    %\path (IOW.north) ++ (-4,0) coordinate (IOWN8);
    
    \draw[very thick,-latex,rounded corners] (GNBP0) to node[right] {$\widetilde{\bm{\lambda}_0}$} (IOWN1);
    \draw[very thick,-latex,rounded corners] (GNBP1) to node[right] {$\widetilde{\bm{\lambda}_1}$} (IOWN3);
    \draw[very thick,-latex,rounded corners] (GNBP5) to node[right] {$\widetilde{\bm{\lambda}_4}$} (IOWN6);
    %\draw[very thick,-latex,rounded corners] (BCS75) -- (IOWN75);
    %\node[below right = 0.35em of BCS75] (RC) {$\bm{\lambda}$};
    %\node[below= 6em of BCS8, rotate=90] (RCLabel) {Residual Connection};
    
    \node [above right=1em of IOWN5] (IWDots) {$\hdots$};
    
    % Systematic Bit Selection
    %\node [draw,line width=.8pt,rounded corners,below= 1.5em of IOW, minimum width=4cm,minimum height=0.6cm] (SBS) {Systematic Bits Selection};
    %\draw[very thick,-latex,rounded corners] (IOW) to node[right] {$\widetilde{\bm{\lambda}}$} (SBS);
    
    %Output and Loss
    \node [below=3em of IOW] (Loss) {$\ell (\widetilde{\mathbf{x}})$};
    \draw[very thick,-latex,rounded corners] (IOW) to node[right] {$\widetilde{\mathbf{x}}$} (Loss);

    %Weights
    \path (BoxRNN) ++ (+6,-2.5) coordinate (WC);
    \node [draw,line width=.8pt,rounded corners,fill=gray!50] (W) at (WC) {$\mathbf{W}^{(n-k),k}$};
    \node [draw,line width=.8pt,rounded corners,above=1.5em of W,text width=2.7cm, align=center] (ConcatH) {Concatenate $\mathbf{W}|\mathbf{I}^{(n-k),(n-k)}$};
    \draw[very thick,-latex,rounded corners] (W) -- (ConcatH);
    \draw[very thick,-latex,rounded corners] (ConcatH) |- (BoxRNN);

    %Training Algo
    \path (Loss) ++ (+3,0) coordinate (LA);
    \node [draw,line width=1.2pt,rounded corners,text width=2cm, align=center, dotted] (LearningAlgo) at (LA) {Learning Algorithm};
    \draw[very thick,-latex,rounded corners, dotted] (Loss) -- (LearningAlgo);
    \draw[very thick,-latex,rounded corners, dotted] (LearningAlgo) -| (W);
    \node [above=24em of ConcatH] (Legend) {\underline{\textbf{Legend:}}};
    \node [draw,line width=.8pt,rounded corners,text width=2.6cm,fill=gray!50, below=0.4em of Legend] (Trainable) {Trainable};
    \node [draw,line width=.8pt,rounded corners,text width=2.6cm,fill=white, below=0.5em of Trainable] (NonTrainable) {Non-trainable};
    %\node (BoxLegend) [draw,line width=.4pt,rounded corners,fit = (Legend) (NonTrainable), scale=1] {};
\end{tikzpicture}}
    \caption{System architecture of the auto-encoder during training. See \cite{GNBP_Decoder} for detailed explanation of the BP RNN Cell architecture and \cite{GNBP_Learning} for comparison with the previous auto-encoder architecture. The number of trainable parameters correspond exactly to the size of the redundant part of the \ac{PC} matrix of the code, \textit{i.e} $(n-k)\times k$.}
    \label{fig:AEarchitecture}
\end{figure}
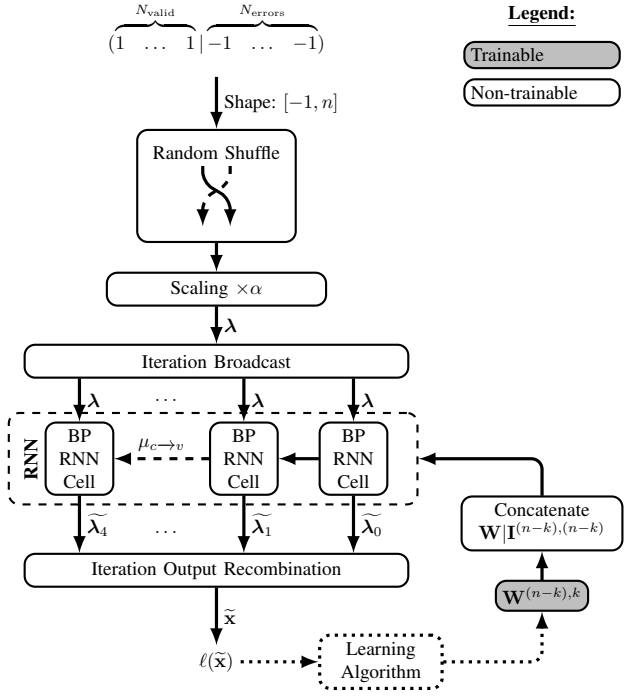

The auto-encoder model proposed for learning error codes is described in Figure~\ref{fig:AEarchitecture}. A trainable \ac{BP} decoder is considered, implementing the standard \ac{BP} algorithm in its forward pass. The differentiable iterative decoding is executed as a \ac{RNN}. The general architecture of the \ac{RNN} cell is based on the one described in \cite{GNBP_Decoder}, without any weighting mechanism within the decoding graph.

The number of BP iterations to evaluate the codes performances is fixed to 5. This choice reflects our focus on applications requiring low computational complexity or real-time operation, as explained in the previous section. While increasing the number of iterations typically improves decoding performance, our primary goal is to construct codes that perform well under these practical efficiency constraints.

The code-word size is $n$ bits, for a code rate of $\frac{k}{n}$. The focus is put here on learning the parity check matrix defining the code, and thus the \ac{BP} factor graph topology. The linearity of the codes and the nature of the decoder enable us to work exclusively with the all-zero code-word during training, while ensuring that the performance of the learned code generalizes to all code-words.

While recent works have explored channel-specific code optimization \cite{DeepCode}, a controlled error channel is employed during the training stage, allowing precise regulation of the number of errors applied to each code-word. We posit that using an \ac{AWGN} channel or a \ac{BSC} instead would not provide any additional relevant information to the optimization procedure, while potentially diminishing our control over the actual learning process. Indeed, to achieve a reasonable error rate for training with AWGN channels, we often need to reduce the noise power to levels where many codewords present little to no challenge for correction, resulting in inefficient training where a significant portion of the dataset contributes minimally to learning. Conversely, a noise power set too high can introduce uncorrectable error patterns that produce contradictory learning signals, potentially causing training regression and unstable convergence. The hypothesis behind our controlled approach is that it ensures almost all codewords in the dataset are meaningful for training, focusing on error patterns that can be solved rather than patterns beyond correction capacity. Based on the previous assumption, the unique code-word of size $n$ provided to the model during training is constructed as the concatenation of $N_{\rm{errors}}$ erroneous bits and $N_{\rm{valid}} = n-N_{\rm{errors}}$ valid bits. The number of $N_{\rm{errors}}$ is considered constant for a given code size during training. The erroneous bits of value $1$ are associated with negative \ac{LLR}s of arbitrary value $-1$. The valid bits of value $0$ are associated with positive \ac{LLR}s of arbitrary value $1$. The objective of the decoder is to retrieve $\mathbf{x}$ as the all-one vector of size $n$.

In the evaluation stage, we focus on the AWGN channel as the standard evaluation benchmark. This choice reflects common practice in communication systems, where signal processing techniques such as channel equalization, scrambling or interleaving are specifically designed to mitigate channel impairments by "whitening" the channel response, making the AWGN model a good approximation of practical channel conditions when reaching the channel decoding stage.

To accommodate diverse error patterns, the first operation applied by the auto-encoder is to randomly shuffle the previously described code-word, leading to code-words with a predetermined number of errors at random indices.

A fixed scaling factor $\alpha$ is then applied, such that the \ac{LLR} vector provided to the \ac{BP} decoder contains random combination of $-\alpha$ and $+\alpha$ \ac{LLR}s. This scaling factor facilitates approximate regulation of the magnitude of values propagated through the model's graph during both forward and backward passes of the training process. We hypothesize that maintaining numerical equilibrium within the computation graph is crucial for successful training outcomes.

The scaled \ac{LLR} vector $\bm{\lambda}$ is then broadcasted according to the desired number of decoding iterations and provided to the \ac{BP} \ac{RNN} cells. Each cell executes one iteration of a standard \ac{BP} algorithm, based on the code's factor graph defined by the trainable parity check matrix, $\mathbf{H}$. As in our previous work, we focus exclusively on systematic codes, where the parity check matrix is represented in standard form:

\begin{equation}
    \mathbf{H} = \left(\mathbf{W}^{(n-k),k}|\mathbf{I}^{(n-k),(n-k)}\right)    
\label{eq:H}
\end{equation}

$\mathbf{H}$ is the result of the concatenation of the trainable part of the code, $\mathbf{W}$, a $(n-k) \times k$ matrix of binary values, with an identity matrix of size $(n-k) \times (n-k)$. The standard form ensures linear independence between the parity-check equations, independently of the trainable part values. We would like to emphasize that elements of $\mathbf{W}$ constitute the only trainable parameters of the system.

The decoded $n$ bits code-word obtained at the end of the last \ac{BP} iteration is used to compute the \ac{BCE} loss function:

\begin{equation}
    \ell(\mathbf{x},\tilde{\mathbf{x}}) = \sum \mathbf{x}\mathrm{log}(\tilde{\mathbf{x}}) + (1 - \mathbf{x})\mathrm{log}(1-\tilde{\mathbf{x}})    
\end{equation}

Since we exclusively work with the zero code-word as input, the loss function can be simplified to: 
\begin{equation}
    \ell(\tilde{\mathbf{x}}) = \sum \mathrm{log}(1-\tilde{\mathbf{x}})    
\end{equation}

\section{A Methodology for Comparing Code Learning Techniques}
\label{section:code_comparison_methodology}

\subsection{On the Performance of Code Learning Techniques}
While code construction would usually be performed offline, the search space complexity remains a fundamental challenge. With $2^{(n-k) \times k}$ possible systematic parity check matrices for a $\mathcal{C}(n,k)$ code, efficient exploration or construction methods are essential as exhaustive search becomes intractable even for modest code lengths.

In this context, a common framework can be identified across state-of-the-art papers employing \ac{ML} for code learning. Typically, the training process iteratively modifies and improves the current version of the code, with the final version emerging after this iterative construction process. This approach aligns with standard \ac{DL} methodologies and many \ac{RL} techniques, which rely on step-by-step optimization of the \ac{NN} model or policy towards a pre-defined objective.

The efficacy of this learning approach has been demonstrated in numerous studies cited in Section~\ref{section:intro}. However, a critical question arises, particularly as most of these works focus on the short to medium length regime. While it is expected that the learning technique should surpass random processes, the mere application of a \ac{ML} algorithm does not inherently provide proof of superiority. As mentioned, these techniques typically rely on iterative code construction, implying that the system processes multiple versions of the code during training.

Consider a \ac{DL} model using gradient descent. At each step, the model would evaluate code performance, compute the gradient of the loss, and update weights to generate a new code. After 100 epochs of 100 steps each, the final code could be viewed as the best among 10,000 codes (at most).

This raises a pertinent question: based on a random search of the code space, what are the probabilities of obtaining a code with equivalent (or better) performance among 10,000 samplings? To demonstrate added value, the training process should at least be superior, in a statistically significant manner, to the random search. Failure to show benefit does not necessarily imply that the learning technique performs a random search, but rather that it is statistically not more efficient than a random process.

We primarily raised this question as part of our effort to develop a robust methodology for evaluating the quality of our code learning techniques.

% It is important to note that these observations may not apply universally to all \ac{ML} approaches in the state-of-the-art, given the diverse methods for learning codes and the significance of optimization system architecture. We primarily raised this question as part of our effort to develop a robust methodology for evaluating the quality of our code learning techniques. Our aim is to establish a framework that can objectively assess the efficacy of various algorithms, taking into account their unique characteristics and optimization strategies. This approach could potentially lead to more meaningful comparisons between different methods and guide future research in the field of code learning.

\subsection{Random Search Model}

Our initial step involves gathering relevant statistics from random code searches. As detailed in Section~\ref{section:system_model}, our study focuses on learning codes based on parity check matrices in standard form, decoded with a classic \ac{BP} algorithm. To efficiently evaluate randomly sampled parity check matrices in standard form at each step, we designed a simple \emph{Random Search} \ac{NN} model. The use of a \ac{NN} here is not associated with \ac{ML} techniques, but rather leverages native support for efficient GPU computing capabilities of the Tensorflow library when available.

A significant challenge in this approach is ensuring a fair comparison between randomly sampled codes and learned ones. The fairness criteria depend on the nature of the code, the system architecture, and the specific \ac{ML} technique used for code learning. To address this challenge in our system, we propose comparing codes of equal size (same $(n,k)$) evaluated using an identical decoder (a standard \ac{BP} decoder with 5 iterations). The 1's density of $\mathbf{W}$, the non-systematic part of the parity check matrix, plays a crucial role in determining expected code performance. If we assume the worst case of a ``learning" process equivalent to a 50\% flipping probability of the bits describing $\mathbf{W}$, then the learning will effectively sample parity check matrices with a density that is, on average, equal to the initialization density. Consequently, we incorporate the 1's density of $\mathbf{W}$ as a parameter in our random code search to enhance the fairness of the comparison.

It is important to note that the density is set as a binomial probability: a density of 30\% does not imply that exactly 30\% of $\mathbf{W}$ is composed of 1s, but rather that each element of $\mathbf{W}$ has a 30\% probability of being 1. This probabilistic approach means that even when fixing a specific density, we are still effectively sampling from the entire space of parity check matrices in standard form. This method allows for a comprehensive exploration of the code space while maintaining statistical consistency in the density of the sampled matrices.

For the chosen code size and 1's density of $\mathbf{W}$, the evaluation process of one random code follows these steps: 
\begin{enumerate} 
    \item Randomly sample a standard form parity check matrix using the selected density. 
    \item Construct the generator matrix $\mathbf{G}$ for the encoder based on the parity check matrix. From the definition of $\mathbf{H}$ in Equation~\ref{eq:H}, we have $\mathbf{G} = \left([\mathbf{W}^{(n-k),k}]^\mathrm{T}|\mathbf{I}^{k,k}\right)$, with $.^\mathrm{T}$ the transpose operator.
    \item Evaluate code performance: set up an end-to-end transmission of random binary words, encoded and transmitted as $+1$, $-1$ on an \ac{AWGN} channel at target $\mathrm{E_b/N_0}$ values. Compute \ac{LLR}s as input for a 5-iterations \ac{BP} decoder.
    \item Compute and store \ac{BLER} values associated with the code, using the Agresti-Coull \cite{BinomialConfidenceInterval} method\footnote{The Agresti-Coull method is a statistical technique used to calculate confidence intervals for binomial proportions, which in this context is applied to estimate BLER values with a specified level of reliability.}. Here we consider \ac{BLER} value reliable when there is a 95\% probability that the true value lies within $\pm10$\% of the estimated \ac{BLER}. Until this bound is not reached, we continue to sample new random binary words. 
\end{enumerate}

In our case, we evaluated 12,800 codes for each of the (32,8), (32,16), (32,24), (64,16), (64,32), and (64,48) code sizes and 18 densities among $\{0.1, 0.15, ..., 0.95\}$). From the gathered \ac{BLER} values, we computed statistics for each $\mathrm{E_b/N_0}$ to obtain mean, standard deviation, minimum, and maximum \ac{BLER}.

As a summary, Table~\ref{tab:best_rand} presents the performance of the code that obtained the best BLER at maximum evaluated $\mathrm{E_b/N_0}$, among all tested densities for each code size (230,400 codes evaluated per size).

\begin{table}[ht] 
\centering
\begin{tabular}{|c|c|c|c|c|c|c|}
\hline
\multirow{2}{*}{$\mathrm{E_b/N_0}$} & \multicolumn{3}{c|}{n = 32} & \multicolumn{3}{c|}{n = 64} \\ \cline{2-7} & k = 8 & k = 16 & k = 24 & k = 16 & k = 32 & k = 48 \\ 
\hline
0 dB & 2.7e-1 & 4.6e-1 & 6.2e-1 & 3.9e-1 & 6.3e-1 & 8.6e-1 \\ 
1 dB & 1.8e-1 & 2.9e-1 & 4.5e-1 & 2.3e-1 & 3.9e-1 & 6.3e-1 \\ 
2 dB & 9.5e-2 & 1.5e-1 & 2.3e-1 & 1.0e-1 & 2.0e-1 & 3.5e-1 \\ 
3 dB & 4.1e-2 & 6.5e-2 & 9.8e-2 & 4.0e-2 & 6.2e-2 & 1.4e-1 \\ 
4 dB & 1.4e-2 & 1.9e-2 & 3.1e-2 & 1.3e-2 & 1.5e-2 & 3.4e-2 \\ 
5 dB & 3.7e-3 & 4.2e-3 & 6.8e-3 & 2.3e-3 & 2.5e-3 & 5.5e-3 \\ 
6 dB & 7.8e-4 & 6.6e-4 & 1.0e-3 & 4.1e-4 & 2.3e-4 & 6.4e-4 \\ 
7 dB & 9.8e-5 & 7.9e-5 & 9.4e-5 & 4.1e-5 & 1.5e-5 & 5.2e-5 \\
\hline
\end{tabular}
\captionsetup{justification=centering}
\caption{BLER of best random codes evaluated at maximum $\mathrm{E_b/N_0}$.}
\label{tab:best_rand}
\end{table}

% \begin{table}[h] 
% \centering
% \begin{tabular}{|c|c|c|c|c|c|c|}
% \hline
% \multirow{2}{*}{$\mathrm{E_b/N_0}$(dB)} & \multicolumn{3}{c|}{n = 32} & \multicolumn{3}{c|}{n = 64} \\ \cline{2-7} & k = 8 & k = 16 & k = 24 & k = 16 & k = 32 & k = 48 \\ 
% \hline
% -6.0 & 2.7e-1 & ---    & ---    & 3.9e-1 & ---    & --- \\ 
% -5.0 & 1.8e-1 & ---    & ---    & 2.3e-1 & ---    & --- \\ 
% -4.0 & 9.5e-2 & ---    & ---    & 1.0e-1 & ---    & --- \\ 
% -3.0 & 4.1e-2 & 4.6e-1 & ---    & 4.0e-2 & 6.3e-1 & --- \\ 
% -2.0 & 1.4e-2 & 2.9e-1 & ---    & 1.3e-2 & 3.9e-1 & --- \\ 
% -1.0 & 3.7e-3 & 1.5e-1 & 6.2e-1 & 2.3e-3 & 2.0e-1 & 8.6e-1 \\ 
% 0.0  & 7.8e-4 & 6.5e-2 & 4.5e-1 & 4.1e-4 & 6.2e-2 & 6.3e-1 \\ 
% 1.0  & 9.8e-5 & 1.9e-2 & 2.3e-1 & 4.1e-5 & 1.5e-2 & 3.5e-1 \\ 
% 2.0  & ---    & 4.2e-3 & 9.8e-2 & ---    & 2.5e-3 & 1.4e-1 \\ 
% 3.0  & ---    & 6.6e-4 & 3.1e-2 & ---    & 2.3e-4 & 3.4e-2 \\ 
% 4.0  & ---    & 7.9e-5 & 6.8e-3 & ---    & 1.5e-5 & 5.5e-3 \\ 
% 5.0  & ---    & ---    & 1.0e-3 & ---    & ---    & 6.4e-4 \\ 
% 6.0  & ---    & ---    & 9.4e-5 & ---    & ---    & 5.2e-5 \\ 
% \hline
% \end{tabular}
% \caption{BLER of best random codes.}
% \label{tab:best_rand}
% \end{table}

For statistical performance comparison, we rely on the estimated \ac{CDF} of the random codes performances for each density and code size. We identify the densities providing the best performances at maximum $\mathrm{E_b/N_0}$ for the following points on the \ac{CDF}: minimum (best performance), first quartile, median, third quartile, and maximum (worst performance). We obtain the Table~\ref{tab:rand_best_density}.

\begin{table}[ht]
\centering
\begin{tabular}{|c|c|c|c|c|c|c|}
\hline
n & \multicolumn{3}{c|}{32} & \multicolumn{3}{c|}{64} \\ 
\hline
k & 8 & 16 & 24 & 16 & 32 & 48 \\ 
\hline
best & 0.30 & 0.30 & 0.40 & 0.20 & 0.15 & 0.25 \\ 
25\% & 0.30 & 0.30 & 0.35 & 0.20 & 0.20 & 0.25 \\ 
50\% & 0.30 & 0.30 & 0.40 & 0.20 & 0.20 & 0.25 \\ 
75\% & 0.30 & 0.30 & 0.45 & 0.20 & 0.20 & 0.30 \\ 
worst & 0.35 & 0.40 & 0.50 & 0.25 & 0.25 & 0.35 \\ 
\hline
\end{tabular}
\captionsetup{justification=centering}
\caption{Densities ordered by random code performances in \ac{BLER} at specific points on the \ac{CDF}.}
\label{tab:rand_best_density}
\end{table}

\begin{figure}[th]
    \centering
    \includegraphics[scale=0.5]{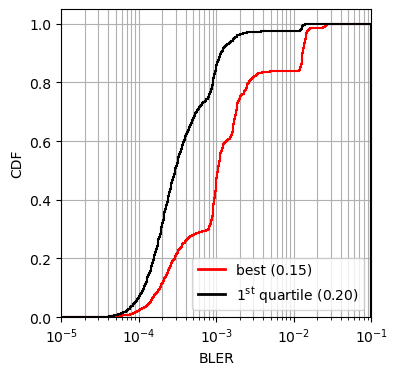}
    \captionsetup{justification=centering}
    \caption{Estimated CDF of \ac{BLER} performance for random codes of size (64,32).}
    \label{fig:cdf_rand}
\end{figure}

As observed, the density providing the best performances and the best 25\% and 50\% of the CDF is usually consistent for a given code size. However, variations due to outliers are evident, as seen in Figure~\ref{fig:cdf_rand}. For (64,32) codes, the best code is obtained with a 15\% density, yet the overall probability of getting a good code is significantly better with a 20\% density. This underscores the importance of examining the CDF, and not just the best code obtained, before conducting statistical comparisons with learned codes.

\subsection{Performance Comparison with Learned Codes}

Having established some random code statistics, we now define a general process to determine whether the studied learning technique surpasses random code search in efficiency. As previously mentioned, for a fairer comparison, we will compare learned codes with random codes of the same size but potentially different densities, acknowledging that \ac{ML} techniques may not maintain the initial 1's density. Based on our previous \ac{CDF} study, we propose to select the density that provides the first quartile performance of the \ac{CDF} for the random codes. The first quartile density constitutes a more conservative benchmark that accounts for variability in random code performance while still representing above-average codes. Moreover, this density is usually also the one providing the best evaluated code.

Once the best configuration of the learning algorithm for the selected code size is defined, the objective is to conduct multiple training sessions, while counting the number of updates performed. The resulting learned code from each training is then associated with its corresponding updates count. For a fair comparison, it is desirable that the total number of updates across all sessions does not exceed the total number of random codes used to generate the statistics.

While this approach may not yield comprehensive statistics on learned codes due to the limited number of resulting codes from each training session, it allows us to assess the relative performance of learned codes against randomly sampled ones. For a training process that has performed M updates, we evaluate the probability of sampling a random matrix that outperforms the resulting learned code, based on M independent random samples, according to the previously estimated CDF.

\subsection{An Example}

\begin{figure}[ht]
    \centering
    \includegraphics[scale=0.55]{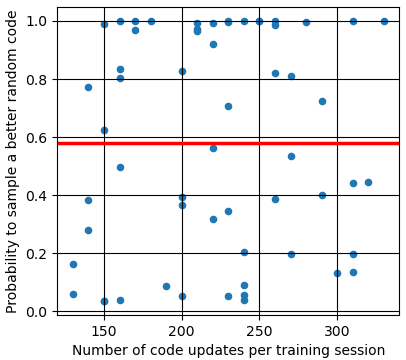}
    \captionsetup{justification=centering}
    \caption{Probability to get a better random code, for size (32,8) and a 1's density of 30\%.}
    \label{fig:RvL}
\end{figure}

Figure~\ref{fig:RvL} illustrates these probabilities for (32,8) codes. The learned codes were obtained using an earlier version of our architecture (not detailed here), and compared with random codes sampled at a 30\% density. Each blue dot represents a learned code, where the x-axis shows the number of updates M during the corresponding training session, and the y-axis indicates the probability of sampling a better code from M random samples. The red line represents the average probability across all learned codes.

From this analysis, we can infer that there is, on average, approximately a 60\% chance of selecting a better code through random search. It is important to note that this does not definitively conclude on the efficacy of the training process or its capacity to produce high-performing codes. However, this visualization serves as a comparative tool for different code learning techniques, with necessary caveats. A technique resulting in a lower red line would indicate greater added value compared to random search.

% This method provides a quantitative basis for comparing learning algorithms, focusing on their ability to consistently produce codes that outperform random selection. While it doesn't capture the full complexity of the learning process, it offers valuable insights into the relative efficiency of different approaches in the context of code generation.

% To further evaluate the efficacy of the learning process, it is valuable to conduct a best-to-best performance comparison. Given that the learning process has examined the same number of codes as the random sampling approach, we can directly compare the top-performing codes from each method. We hypothesize that a well-designed ML algorithm should construct a superior code compared to the best randomly sampled one.

% However, the learning-based campaign takes significantly less time to complete than the random codes ones. This motivates a discussion on the performance vs number of updates (matrices) evaluations.

% So instead of comparing performance just based on best available code, the community could also use this methodology to demonstrate that the learning process is efficient.

\section{Quantization Mechanisms}
\label{section:gradient_quantificatin_mechanisms}

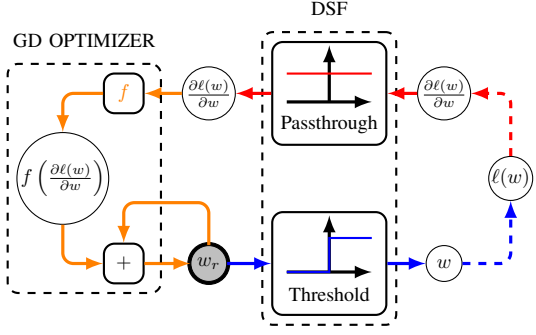
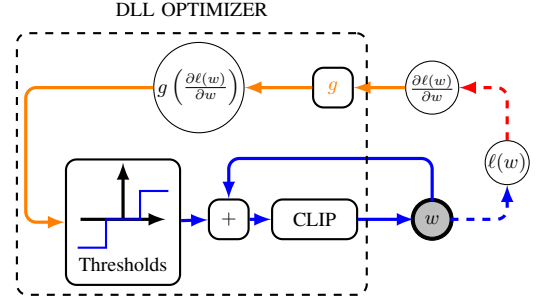
\begin{figure*}
    \begin{subfigure}[b]{0.48\textwidth}
        \centering
        \begin{tikzpicture}[scale=0.75,every node/.style={transform shape} ] %every node/.style={scale=0.9}]
    \contourlength{1.4pt}
    \tikzset{>=latex} % for LaTeX arrow head
    %Manually reduce line spacing to single space no matter double or single column document (draft mode)
    \linespread{1}

    % STEP
    \path (0,0) ++ (+0,0) coordinate (STEP1);
    \path (STEP1) ++ (+0.75,0) coordinate (STEP2);
    \path (STEP2) ++ (+0.75,0) coordinate (STEP3);
    \path (STEP2) ++ (+0,1) coordinate (STEP4);
    \path (STEP2) ++ (+0,0.6) coordinate (STEP5);
    \path (STEP3) ++ (+0,0.6) coordinate (STEP6);
    \draw[very thick,-latex] (STEP1) -- (STEP3);
    \draw[very thick,-latex] (STEP2) -- (STEP4);
    \draw[thick,blue] (STEP1) -- (STEP2) -- (STEP5) -- (STEP6);

    \path (STEP1) ++ (-0.1,-0.6) coordinate (STEPBOX1);
    \path (STEP3) ++ (+0.1,+0.9) coordinate (STEPBOX2);
    \node (STEP) [draw,line width=.8pt,rounded corners,fit = (STEPBOX1) (STEPBOX2)] {};
    \path (STEP2) ++ (0,-0.4) coordinate (STEPLABEL1);
    \node [align=center] (STEPLABEL) at (STEPLABEL1) {Threshold};

    % W
    \path (STEP.west) ++ (-1.1,0) coordinate (W1);
    \node [circle,minimum size=0.1cm,inner sep=0.1,draw,minimum size=21pt,fill=black] (W) at (W1) {$w_r$};
    \node [circle,minimum size=0.1cm,inner sep=0.1,draw,minimum size=18pt,fill=lightgray] (W) at (W1) {$w_r$};
    
    % f(W)
    \path (STEP.east) ++ (+1,0) coordinate (F1);
    \node [circle,minimum size=0.1cm,inner sep=0.1,draw,minimum size=18pt] (F) at (F1) {$w$};

    % PASSTHROUGH GRAD
    \path (STEP1) ++ (0,+3) coordinate (PASS1);
    \path (PASS1) ++ (+0.75,0) coordinate (PASS2);
    \path (PASS2) ++ (+0.75,0) coordinate (PASS3);
    \path (PASS2) ++ (+0,1) coordinate (PASS4);
    \path (PASS1) ++ (+0,0.5) coordinate (PASS5);
    \path (PASS3) ++ (+0,0.5) coordinate (PASS6);
    \draw[very thick,-latex] (PASS1) -- (PASS3);
    \draw[very thick,-latex] (PASS2) -- (PASS4);
    \draw[thick,red] (PASS5) -- (PASS6);

    \path (PASS1) ++ (-0.1,-0.6) coordinate (PASSBOX1);
    \path (PASS3) ++ (+0.1,+0.9) coordinate (PASSBOX2);
    \node (PASS) [draw,line width=.8pt,rounded corners,fit = (PASSBOX1) (PASSBOX2)] {};
    \path (PASS2) ++ (0,-0.4) coordinate (PASSLABEL1);
    \node [align=center] (PASSLABEL) at (PASSLABEL1) {Passthrough};

    % DSF BOX
    \node (DSF) [draw,line width=.8pt,rounded corners,dashed,fit = (PASS) (STEP)] {};
    \path (DSF.north) ++ (0,+0.4) coordinate (DSFLABEL1);
    \node [align=center] (DSFLABEL) at (DSFLABEL1) {DSF};

     % l(W)
    \path (DSF.east) ++ (+2,0) coordinate (L1);
    \node [circle,minimum size=0.1cm,inner sep=0.1,draw,minimum size=18pt] (L) at (L1) {$\ell(w)$};

    % df(W)/dl(w_r)
    \path (PASS.east) ++ (+1,+0) coordinate (DLDF1);
    \node [circle,minimum size=0.1cm,inner sep=0.1,draw,minimum size=18pt] (DLDF) at (DLDF1) {$\frac{\partial \ell(w)}{\partial w}$};

    % ~df(W)/dl(w_r)
    \path (PASS.west) ++ (-1.1,+0) coordinate (DLDW1);
    \node [circle,minimum size=0.1cm,inner sep=0.1,draw,minimum size=18pt] (DLDW) at (DLDW1) {$\frac{\partial \ell(w)}{\partial w}$};

    % ALPHA
    \path (DLDW) ++ (-1.5,0) coordinate (ID1);
    % \path (ID1) ++ (+0.75,0) coordinate (ID2);
    % \path (ID2) ++ (+0.75,0) coordinate (ID3);
    % \path (ID2) ++ (+0,1) coordinate (ID4);
    % \path (ID1) ++ (+0,+0.5) coordinate (ID5);
    % \path (ID3) ++ (+0,-0.5) coordinate (ID6);
    % \draw[very thick,-latex] (ID1) -- (ID3);
    % \draw[very thick,-latex] (ID2) -- (ID4);
    % \draw[thick,orange] (ID5) -- (ID6);

    % \path (ID1) ++ (-0.2,-0.4) coordinate (IDBOX1);
    % \path (ID1) ++ (+0.2,+0.4) coordinate (IDBOX2);
    \node [draw,line width=.8pt,rounded corners, minimum size =0.7cm] (ID) at (ID1) {\color{orange}$f$};
    % \path (ID2) ++ (0,-0.4) coordinate (IDLABEL1);
    % \node [align=center] (IDLABEL) at (IDLABEL1) {$-\alpha$};

    % A~df(W)/dl(w_r)
    \path (DSF.west) ++ (-3.5,0) coordinate (ADLDW1);
    \node [circle,minimum size=0.1cm,inner sep=0.1,draw,minimum size=18pt] (ADLDW) at (ADLDW1) {$f\left(\frac{\partial \ell(w)}{\partial w}\right)$};

    % ADD
    \path (W) ++ (-1.5,0) coordinate (ADD1);
    \node [draw,minimum size=0.7cm,inner sep=0.1,line width=.8pt,rounded corners] (ADD) at (ADD1) {$+$};

    % CONNECT
    \draw[very thick,-latex,rounded corners,blue] (W) -- (STEP);  
    \draw[very thick,-latex,rounded corners,blue] (STEP) -- (F);
    \draw[very thick,-latex,rounded corners,dashed,blue] (F) -| (L); 
    \draw[very thick,-latex,rounded corners,dashed,red] (L) |- (DLDF);
    \draw[very thick,-latex,rounded corners,red] (DLDF) -- (PASS);
    \draw[very thick,-latex,rounded corners,red] (PASS) -- (DLDW);
    \draw[very thick,-latex,rounded corners,orange] (DLDW) -- (ID);
    \draw[very thick,-latex,rounded corners,orange] (ID) -| (ADLDW);
    \draw[very thick,-latex,rounded corners,orange] (ADLDW) |- (ADD);
    \draw[very thick,-latex,rounded corners,orange] (ADD) -- (W);

    \path (W.north) ++ (-1,+0.75) coordinate (X);
    \draw[very thick,-latex,rounded corners,orange] (W) |- (X) -| (ADD);
    
    % OPTIMIZER
    % \path (ID.north) ++ (+1.2,+0) coordinate (OPTBOX1);
    % \path (ADLDW.south) ++ (-0.8,-1) coordinate (OPTBOX2);
    \node (OPT) [draw,line width=.8pt,rounded corners, dashed, fit = (ADD)(ADLDW)(ID)(X)] {};
    \path (OPT.north) ++ (0,+0.4) coordinate (OPTLABEL1);
    \node [align=center] (OPTLABEL) at (OPTLABEL1) {GD OPTIMIZER};
    
\end{tikzpicture}
        \captionsetup{justification=centering}
        \caption{A conventional Gradient Descent optimizer is associated with a specific graph.}
        \label{fig:GD}
    \end{subfigure}
    \hfill
    \begin{subfigure}[b]{0.48\textwidth}
        \centering
        \begin{tikzpicture}[scale=0.75,every node/.style={transform shape} ] %every node/.style={scale=0.9}]
    \contourlength{1.4pt}
    \tikzset{>=latex} % for LaTeX arrow head
    %Manually reduce line spacing to single space no matter double or single column document (draft mode)
    \linespread{1}
    \node [circle,minimum size=21pt,inner sep=0.1,draw,fill=black] (W) at (0,0) {$w$};
    \node [circle,minimum size=0.1cm,inner sep=0.1,draw,minimum size=18pt,fill=lightgray] (W) at (0,0) {$w$};

     % l(W)
    \path (W.east) ++ (+1,1) coordinate (L1);
    \node [circle,minimum size=0.1cm,inner sep=0.1,draw,minimum size=18pt] (L) at (L1) {$\ell(w)$};
    
    % ~df(W)/dl(w_r)
    \path (W.north) ++ (0,+2) coordinate (DLDW1);
    \node [circle,minimum size=0.1cm,inner sep=0.1,draw,minimum size=18pt] (DLDW) at (DLDW1) {${\frac{\partial \ell(w)}{\partial w}}$};

    % ALPHA
    \path (DLDW) ++ (-1.75,0) coordinate (ID1);
    % \path (ID1) ++ (+0.75,0) coordinate (ID2);
    % \path (ID2) ++ (+0.75,0) coordinate (ID3);
    % \path (ID2) ++ (+0,1) coordinate (ID4);
    % \path (ID1) ++ (+0,+0.5) coordinate (ID5);
    % \path (ID3) ++ (+0,-0.5) coordinate (ID6);
    % \draw[very thick,-latex] (ID1) -- (ID3);
    % \draw[very thick,-latex] (ID2) -- (ID4);
    % \draw[thick,orange] (ID5) -- (ID6);

    % \path (ID1) ++ (-0.2,-0.4) coordinate (IDBOX1);
    % \path (ID1) ++ (+0.2,+0.4) coordinate (IDBOX2);
    \node [draw,line width=.8pt,rounded corners, minimum size=0.7cm] (ID) at (ID1) {\color{orange}$g$};

    % % ~alphadl(w_r)/dw
    % \path (ID.west) ++ (-1.5,0) coordinate (ADLDW1);
    % \node [circle,minimum size=0.1cm,inner sep=0.1,draw,minimum size=18pt] (ADLDW) at (ADLDW1) {${-\alpha\frac{\partial \ell(w)}{\partial w}}$};

    % % ADD
    % \path (ADLDW.west) ++ (-1.5,0) coordinate (ADD1);
    % \node [draw,minimum size=0.7cm,inner sep=0.1,line width=.8pt,rounded corners] (ADD) at (ADD1) {$+$};

    % g dl(w_r)/dw
    \path (ID.west) ++ (-2,0) coordinate (ASDLDW1);
    \node [circle,minimum size=0.1cm,inner sep=0.1,draw,minimum size=18pt] (ASDLDW) at (ASDLDW1) {$g\left(\frac{\partial \ell(w)}{\partial w}\right)$};

    % CLIP
    \path (W.west) ++ (-1.75,0) coordinate (CLIP1);
    \node [draw,minimum size=0.7cm,minimum width=1.5cm,inner sep=0.1,line width=.8pt,rounded corners] (CLIP) at (CLIP1) {CLIP};

    % ADD
    \path (CLIP.west) ++ (-0.75,0) coordinate (ADDW1);
    \node [draw,minimum size=0.7cm,inner sep=0.1,line width=.8pt,rounded corners] (ADDW) at (ADDW1) {$+$};

    % STEP
    \path (ADDW.west) ++ (-2.25,0) coordinate (STEP1);
    \path (STEP1) ++ (+0.75,0) coordinate (STEP2);
    \path (STEP2) ++ (+0.75,0) coordinate (STEP3);
    \path (STEP2) ++ (+0,1) coordinate (STEP4);
    
    \path (STEP2) ++ (+0.3,0) coordinate (STEP5);
    \path (STEP5) ++ (+0,0.5) coordinate (STEP6);
    \path (STEP6) ++ (+0.5,0) coordinate (STEP7);

    \path (STEP2) ++ (-0.3,0) coordinate (STEP8);
    \path (STEP8) ++ (+0,-0.5) coordinate (STEP9);
    \path (STEP9) ++ (-0.5,0) coordinate (STEP10);
    
    \draw[very thick,-latex] (STEP1) -- (STEP3);
    \draw[very thick,-latex] (STEP2) -- (STEP4);
    \draw[thick,blue] (STEP10) -- (STEP9) -- (STEP8) -- (STEP5)  -- (STEP6)  -- (STEP7);

    \path (STEP1) ++ (-0.1,-1) coordinate (STEPBOX1);
    \path (STEP3) ++ (+0.1,+0.9) coordinate (STEPBOX2);
    \node (STEP) [draw,line width=.8pt,rounded corners,fit = (STEPBOX1) (STEPBOX2)] {};
    \path (STEP2) ++ (0,-0.8) coordinate (STEPLABEL1);
    \node [align=center] (STEPLABEL) at (STEPLABEL1) {Thresholds};

    % CONNECT
    \draw[very thick,-latex,rounded corners, dashed, blue] (W) -| (L);  
    \draw[very thick,-latex,rounded corners,dashed, red] (L) |- (DLDW);
    \draw[very thick,-latex,rounded corners,orange] (DLDW) -- (ID); 
    % \draw[very thick,-latex,rounded corners,orange] (ID) -- (ADLDW);
    % \draw[very thick,-latex,rounded corners,orange] (ADLDW) -- (ADD);
    \draw[very thick,-latex,rounded corners,orange] (ID) -- (ASDLDW);
    \path (ASDLDW.west) ++ (-2.25,-0.75) coordinate (X1);
    \draw[very thick,-latex,rounded corners,orange] (ASDLDW) -| (X1) |- (STEP);
    \draw[very thick,-latex,rounded corners,blue] (STEP) -- (ADDW);
    \draw[very thick,-latex,rounded corners,blue] (ADDW) -- (CLIP);
    \draw[very thick,-latex,rounded corners,blue] (CLIP) -- (W);

    \path (W.north) ++ (-2,+0.75) coordinate (X2);
    \draw[very thick,-latex,rounded corners,blue] (W) |- (X2) -| (ADDW);
    %\draw[very thick,-latex,rounded corners,orange] (ASDLDW) -| (ADD);

    % OPTIMIZER
    %\path (ID.north) ++ (+1,+0) coordinate (OPTBOX1);
    \node (OPT) [draw,line width=.8pt,rounded corners, dashed, fit = (CLIP)(STEP)(X1)(ID)(ASDLDW)] {};
    \path (OPT.north) ++ (0,+0.4) coordinate (OPTLABEL1);
    \node [align=center] (OPTLABEL) at (OPTLABEL1) {DLL OPTIMIZER};
\end{tikzpicture}
        \captionsetup{justification=centering}
        \caption{A specialized optimizer relying on gradient quantization is associated with a standard graph.}
        \label{fig:DLL}
    \end{subfigure}
    \captionsetup{justification=centering}
    \caption{Two classes of method to learn discrete-valued weights.
    Trainable weights are represented in gray. Related operations in both figures are depicted with the same color.}
    \label{fig:GD_vs_DLL}
\end{figure*}

The optimization of trainable weights in \ac{NN}s typically relies on gradient descent algorithm and its variants. These techniques necessitate a differentiable computational graph for backpropagation, preventing the direct incorporation of discrete functions and values. This constraint poses a significant challenge for learning the parity check matrix, whose elements are inherently binary. To address this challenge, state-of-the-art approaches \cite{GNBP_Learning,DL-Polar}, \cite{Shlezinger2022} employ specialized approximation of non-differentiable function \cite{DAB_ramapuram}. Such function quantizes the real-valued numbers to binary values during the forward pass, while approximating the function's gradient in the backward pass to ensure differentiability. Such \ac{DSF} can be defined as (here as a direct pass-through):

\begin{equation} 
\label{eq:dsf}
\left\{ 
\begin{aligned} 
f(x) &= \text{step}(x) \\ 
\frac{df(x)}{dx} &= 1
\end{aligned} \right. 
\end{equation}

Nevertheless, it is worth considering the use of scalar weights that can only assume values of 0 or 1, behaving virtually as binary weights. In this scenario, considering a continuous extension between these two values, it is possible to compute the gradient. Consequently, contrarily to the aforementioned \ac{DSF}, there is no need to use an approximate gradient within the \ac{NN} model's graph. However, conventional gradient descent algorithms would cause the scalar weights to deviate from their constrained binary values 0 and 1. Therefore, we must design a dedicated optimizer to maintain these discrete values during training. We denote this class of optimizers as \ac{DLL} optimizers. This binary approach could provide a more interpretable framework, as both the weights and their associated discrete updates naturally align with various discrete problems, including error correction. Consequently, it could enhance our understanding of the optimization process and potentially offer finer control over the optimizer's behavior. 

Figure~\ref{fig:GD_vs_DLL} illustrates these two distinct approaches. The DSF method, from the literature (and used in our previous work), typically employs a conventional gradient descent optimizer, gradually modifying the real-valued trainable weights through accumulated updates. The quantization mechanism ensuring binary compatibility is integrated into the neural network itself, implemented through a specialized forward and backward computational graph. In contrast, the proposed DLL approach utilizes a standard neural network graph but incorporates a specialized optimizer. This optimizer quantizes the gradient to update the real weights directly to discrete values, preserving their binary nature throughout the training process. It is worth noting that under certain conditions and configurations, these two approaches to the same problem, although conceptually different, can exhibit certain equivalences, e.g. instead of accumulating small update on the real parameters, one could envision to accumulate gradients.

As a general definition, we consider a discrete trainable weight denoted by $w$, which takes its values from a discrete ordered set $S$. The elements in $S$ are assumed to be real numbers. In many gradient-based optimizers, the update step for the weight value takes the following general form:

% As explained, the discrete nature of the weights primarily stems from the design of the automatic updates applied to them, effectively maintaining the system within a finite number of operating points. Despite their discrete nature, these weights are treated as continuous scalar values within the computational graph. This approach ensures that the graph remains differentiable, allowing for the computation of gradients with respect to a defined loss function.

\begin{equation}
    w \gets w + g(\nabla_w\ell,\boldsymbol{\phi},w)
\end{equation}

where $g$ denotes the optimizer function, $\nabla_w\ell$ represents the gradient of the loss function $\ell$ with respect to the weight $w$ and $\boldsymbol{\phi}$ is a set of hyper-parameters.

In the context of discrete weights, the function $g$ should output values that guide the transition of the weight within its discrete set $S$. Given that the set $S$ is ordered, it is intuitive to consider that the weight value should evolve in the opposite direction of the gradient, analogous to traditional gradient descent algorithms. The remaining variable to determine is the magnitude of the step to take. In the case of binary weights, the problem simplifies: the update value can only be $-1$, $+1$, or $0$.

\subsection{Gradient Quantization LeArning Techniques}

In the following, we propose a method to quantize the scalar gradient value into discrete updates for the trainable weights. We term this approach Gradient Quantization LeArning (GQLA), as part of DLL optimizers. The quantization operation necessitates defining thresholds that map the current gradient to possible update values. Determining these thresholds can be challenging, as they depend on the expected gradient magnitudes, the elements in $S$, and more broadly, on the impact of discrete weight changes on the rest of the optimization process. Moreover, each discrete weight could potentially have its own set of thresholds, as well as its own set $S$, increasing the complexity of the problem.

\subsubsection{Mini-Batch GQLA}

Even in our system, where binary weights necessitate the definition of only two thresholds (one positive and one negative), this task remains difficult. The variations in error combinations in code-words across different batches, as well as the considered code-rate itself, affect the gradient magnitudes in a manner that we find difficult to interpret consistently. This is further compounded by the fact that the range of gradient values 
constantly evolve during the training. Consequently, we have chosen to rely solely on the sign of the gradient, independently of its actual numerical value.

Based on this approach, we propose a straightforward optimizer behavior as follows:

\begin{enumerate} 
    \item If the gradient is positive, the weight should be set to zero. The update is then $-1$ if $w=1$, or $0$ if $w=0$. 
    \item If the gradient is negative, the weight should be set to one. The update is then $+1$ if $w=0$, or $0$ if $w=1$. 
    \item If the gradient is zero, the update should be $0$. 
\end{enumerate}

This first algorithm, the Mini-Batch GQLA (MB-GQLA) for binary weight, is described in Algorithm~\ref{alg:mb_gqla}. The matrix $\mathbf{Q}$ that stores the gradients' signs, has the same size as $\mathbf{W}$, with $(i,j)$ the index of a single element in the 2D-matrix. Using this algorithm as our (DLL) optimizer, we are able to employ trainable binary weights in the parity check matrix, while relying on a traditional gradient computation.

However, conventional gradient descent algorithms are generally effective in traditional \ac{NN}s thanks to the small incremental weights' updates that navigate smoothly the loss slope. The discrete nature of our optimizer introduces discontinuities in the exploration of the loss landscape. This is especially true in our system, where a single bit flip can dramatically alter the code's structure and performance. Based on the current MB-GQLA description, many bits may flip at each batch. Consequently, we require a technique to enhance our confidence in the bit flipping decisions.

\begin{algorithm}
\caption{Mini-Batch GQLA}
\label{alg:mb_gqla}
\begin{algorithmic}[1]
\Require Binary weights matrix $\mathbf{W}$.
\Ensure Updated weights matrix $\mathbf{W}$.
\For {batch $k$}
    \State Compute gradient $\nabla_\mathbf{W}\ell$ of loss function $\ell$ with respect to the weights $\mathbf{W}$.
    \State $\mathbf{Q} \gets \text{sign}(\nabla_\mathbf{W}\ell)$
    \If{($\mathbf{Q}_{(i,j)} = +1$ and $\mathbf{W}_{(i,j)} = 0$) or ($\mathbf{Q}_{(i,j)} = -1$ and \phantom{m} $\mathbf{W}_{(i,j)} = 1$)}
        \State $\mathbf{W}_{(i,j)} \gets \mathbf{W}_{(i,j)} + 0$
    \Else
        \State $\mathbf{W}_{(i,j)} \gets \mathbf{W}_{(i,j)} - \mathbf{Q}_{(i,j)}$
    \EndIf
\EndFor
\end{algorithmic}
\end{algorithm}

\subsubsection{Update Matrix}

We propose using a dedicated Update Matrix to store bit flipping decisions for each batch. This matrix, denoted $\mathbf{U}$, has the same dimensions as $\mathbf{W}$ and functions similarly to a matrix of counters. Specifically, elements of $\mathbf{U}$ accumulates the computed gradient sign value for each weight as $-1$, $0$, or $+1$, until a stopping criterion is met, effectively triggering the update of the weights. One of the aforementioned challenges of gradient quantization is to define thresholds based on gradient magnitude. The Update Matrix provides a tool to define discrete thresholds based on gradient sign accumulation, independent of gradient magnitude. The more consistently the gradient indicates a specific update direction for a weight across multiple batches, the higher our confidence in applying that update. This behavior can be seen as analogous to momentum-based optimizers like ADAM.

Building upon this idea, we define a single absolute threshold value, denoted T. When at least one accumulator reaches the value T in absolute, all weights whose accumulators equal T are updated. The remaining weights retain their current values. The Update Matrix is then reset, \emph{i.e.} all accumulator values are set to zero. Indeed, once some bits are flipped, the code is modified, and the next accumulation process should not be influenced by the previous code. This learning algorithm is summarized in Algorithm~\ref{alg:mb_gqla_um}. It is worth noting that the threshold T remains constant throughout the process. Consequently, the trigger of an update does not depend on the number of batches processed, but rather on the predominant update direction of at least one accumulator exceeding this fixed threshold. This approach ensures that updates are driven by consistent directional signals.

% The combination of the MB-GQLA algorithm with the Update Matrix addresses the challenges mentioned earlier for learning discrete weights. 
% To further enhance this approach, we consider the individual contributions of samples within each batch. We propose an additional refinement to our algorithm that ensures each sample exerts equal influence on the accumulated batch sign.

\begin{algorithm}
\caption{Mini-Batch GQLA with Update Matrix}
\label{alg:mb_gqla_um}
\begin{algorithmic}[1]
\Require Binary weights matrix $\mathbf{W}$ and associated Update Matrix $\mathbf{U}$, threshold value $\rm{T}$.
\Ensure Updated weights $\mathbf{W}$.
\State $\mathbf{U} \gets 0$
\For {batch $k$}
    \State Compute gradient $\nabla_\mathbf{W}\ell$ of loss function $\ell$ with respect to the weights $\mathbf{W}$.
    \State $\mathbf{U} \gets \mathbf{U} + \text{sign}(\nabla_\mathbf{W}\ell)$
    \If {${\text{max}(\lvert\mathbf{U}\rvert) = \rm{T}}$}
        \ForAll {$\lvert\mathbf{U}_{(i,j)}\rvert = \rm{T}$}
            \If{($\mathbf{U}_{(i,j)} = +\rm{T}$ and $\mathbf{W}_{(i,j)} = 0$) or \phantom{mm mm} \phantom{m} ($\mathbf{U}_{(i,j)} = -\rm{T}$ and $\mathbf{W}_{(i,j)} = 1$)}
                \State $\mathbf{W}_{(i,j)} \gets \mathbf{W}_{(i,j)} + 0$
            \Else
                \State $\mathbf{W}_{(i,j)} \gets \mathbf{W}_{(i,j)} - \frac{\mathbf{U}_{(i,j)}}{\rm{T}}$
            \EndIf
        \EndFor
        \State $\mathbf{U} \gets 0$
    \EndIf
\EndFor
\end{algorithmic}
\end{algorithm}

\subsubsection{Stochastic GQLA}

We extend the GQLA concept to the sample level, computing the gradient of the weights with respect to each sample in the batch. This approach is equivalent to calculating the Jacobian of the loss function before any reduction operation on the batch dimension, or to employing the \ac{SGD} algorithm. We designate this algorithm as Stochastic GQLA (S-GQLA), deriving its name from this stochastic characteristic.

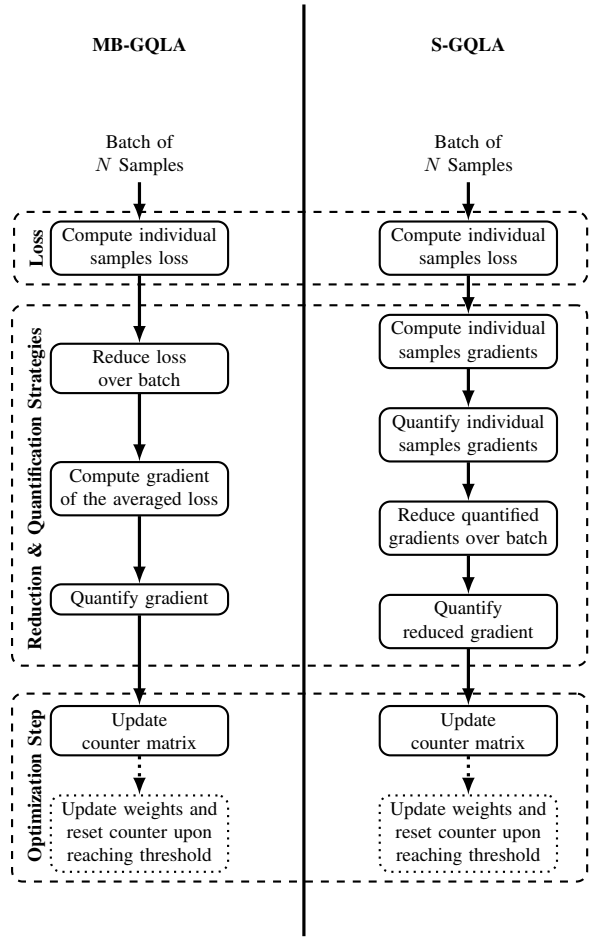
\begin{figure}[htbp]
    \centerline{\usetikzlibrary{fit,backgrounds}
\contourlength{1.4pt}
\tikzset{>=latex} % for LaTeX arrow head

\begin{tikzpicture}[scale=0.725,every node/.style={transform shape} ] %every node/.style={scale=0.9}]
    %Manually reduce line spacing to single space no matter double or single column document (draft mode)
    \linespread{1}
    %
    
    % Input
    \node (Origin) at (0,0) {};
    \path (Origin) ++ (+0,+0) coordinate (O);
    \path (O) ++ (+0,-17) coordinate (E);

    \draw[very thick,rounded corners] (O) -- (E);

    \path (O) ++ (+3,-0.75) coordinate (R);
    \node [text width=3.5cm, align=center] (NewAlgorithm) at (R) {\textbf{S-GQLA}};
    \path (R) ++ (0,-2) coordinate (R1);
    \node [text width=2cm, align=center] (RBatchSamples) at (R1) {Batch of $N$ Samples};
    \node [draw,line width=.8pt,rounded corners, text width=3cm, align=center, below=2em of RBatchSamples] (RLoss) {Compute individual samples loss};
    \node [draw,line width=.8pt,rounded corners, text width=3cm, align=center, below=2em of RLoss] (RGradient) {Compute individual samples gradients};
    \node [draw,line width=.8pt,rounded corners, text width=3cm, align=center, below=2em of RGradient] (RQuantification1) {Quantify individual samples gradients};    
    \node [draw,line width=.8pt,rounded corners, text width=3cm, align=center, below=2em of RQuantification1] (RReduction) {Reduce quantified gradients over batch};
    \node [draw,line width=.8pt,rounded corners, text width=3cm, align=center, below=2em of RReduction] (RQuantification2) {Quantify reduced gradient}; 
    \path (R1) ++ (0,-10.5) coordinate (R2);
    \node [draw,line width=.8pt,rounded corners, text width=3cm, align=center] (RUpdate) at (R2) {Update counter matrix };  
    \node [draw,line width=.8pt,rounded corners, text width=3cm, align=center, below=2em of RUpdate, dotted] (RUpdateWeights) {Update weights and reset counter upon reaching threshold};  

    \draw[very thick,-latex,rounded corners] (RBatchSamples) -- (RLoss);
    \draw[very thick,-latex,rounded corners] (RLoss) -- (RGradient);
    \draw[very thick,-latex,rounded corners] (RGradient) -- (RQuantification1);
    \draw[very thick,-latex,rounded corners] (RQuantification1) -- (RReduction);
    \draw[very thick,-latex,rounded corners] (RReduction) -- (RQuantification2);
    \draw[very thick,-latex,rounded corners] (RQuantification2) -- (RUpdate);
    \draw[very thick,-latex,rounded corners, dotted] (RUpdate) -- (RUpdateWeights);

    \path (O) ++ (-3,-0.75) coordinate (L);
    \node [text width=3.5cm, align=center] (MiniBatch) at (L) {\textbf{MB-GQLA}};
    \path (L) ++ (0,-2) coordinate (L1);
    \node [text width=2cm, align=center] (LBatchSamples) at (L1) {Batch of $N$ Samples};
    \node [draw,line width=.8pt,rounded corners, text width=3cm, align=center, below=2em of LBatchSamples] (LLoss) {Compute individual samples loss};
    \node [draw,line width=.8pt,rounded corners, text width=3cm, align=center, below=3.5em of LLoss] (LReduction) {Reduce loss over batch};
    \node [draw,line width=.8pt,rounded corners, text width=3cm, align=center, below=3.5em of LReduction] (LGradient) {Compute gradient of the averaged loss};
    \node [draw,line width=.8pt,rounded corners, text width=3cm, align=center, below=3.5em of LGradient] (LQuantification1) {Quantify gradient}; 
    \path (L1) ++ (0,-10.5) coordinate (L2);
    \node [draw,line width=.8pt,rounded corners, text width=3cm, align=center] (LUpdate) at (L2) {Update counter matrix };  
    \node [draw,line width=.8pt,rounded corners, text width=3cm, align=center, below=2em of LUpdate, dotted] (LUpdateWeights) {Update weights and reset counter upon reaching threshold};  

    \draw[very thick,-latex,rounded corners] (LBatchSamples) -- (LLoss);
    \draw[very thick,-latex,rounded corners] (LLoss) -- (LReduction);
    \draw[very thick,-latex,rounded corners] (LReduction) -- (LGradient);    
    \draw[very thick,-latex,rounded corners] (LGradient) -- (LQuantification1);
    \draw[very thick,-latex,rounded corners] (LQuantification1) -- (LUpdate);
    \draw[very thick,-latex,rounded corners, dotted] (LUpdate) -- (LUpdateWeights);

    \path (LLoss.west) ++ (-0.25,+0) coordinate (LabelLossC);
    \path (RLoss.east) ++ (+0.25,+0) coordinate (LabelLossCPhantom);
    \node [align=center, rotate=90] (LabelLoss) at (LabelLossC) {\textbf{Loss}};
    \node [align=center, rotate=90] (LabelLossPhantom) at (LabelLossCPhantom) {\textbf{}};
    \node (BoxQuantificationStrategy) [draw,line width=.8pt,dashed, rounded corners,fit = (LLoss) (RLoss) (LabelLoss) (LabelLossPhantom), scale=1] {};
    
    \path (LGradient.west) ++ (-0.25,+0) coordinate (LabelC);
    \path (RGradient.east) ++ (+0.25,+0) coordinate (LabelCPhantom);
    \node [align=center, rotate=90] (Label) at (LabelC) {\textbf{Reduction \& Quantification Strategies}};
    \node [align=center, rotate=90] (LabelPhantom) at (LabelCPhantom) {\textbf{}};
    \node (BoxQuantificationStrategy) [draw,line width=.8pt,dashed, rounded corners,fit = (LReduction) (RGradient) (RQuantification2) (Label) (LabelPhantom), scale=1] {};

    \path (LUpdate.west) ++ (-0.25,-1) coordinate (LabelOptimizerC);
    \path (RUpdate.east) ++ (+0.25,+0) coordinate (LabelOptimizerCPhantom);
    \node [align=center, rotate=90] (LabelOptimizer) at (LabelOptimizerC) {\textbf{Optimization Step}};
    \node [align=center, rotate=90] (LabelOptimizerPhantom) at (LabelOptimizerCPhantom) {\textbf{}};
    \node (BoxOptimizer) [draw,line width=.8pt,dashed, rounded corners,fit = (LUpdate) (RUpdateWeights) (LabelOptimizer) (LabelOptimizerPhantom), scale=1] {};
    
\end{tikzpicture}}
    \captionsetup{justification=centering}
    \caption{Description of MB-GQLA and S-GQLA algorithms associated with the Update Matrix.}
    \label{fig:LearningComp}
\end{figure}

The S-GQLA considers the sign of each individual gradient. These signs are summed over a batch, and we then take the sign of the result to obtain values in $\{+1, -1, 0\}$ for each trainable weight. Consequently, each sample in the batch has the same impact on the accumulator increment in the Update Matrix. The complete algorithm is described in Algorithm~\ref{alg:s_gqla}. Figure~\ref{fig:LearningComp} illustrates the complete MB-GQLA and S-GQLA algorithms in conjunction with the Update Matrix. 

\begin{algorithm}
\caption{Stochastic GQLA with Update Matrix}
\label{alg:s_gqla}
\begin{algorithmic}[1]
\Require Binary weights matrix $\mathbf{W}$ and associated Update Matrix $\mathbf{U}$, threshold value $\rm{T}$.
\Ensure Updated weights $\mathbf{W}$.
\State Update matrix $\mathbf{U} \gets 0$
\For {batch $k$}
    \State Quantization matrix $\mathbf{Q} \gets 0$
    \For {sample $s$ in batch $k$}
        \State Compute gradient $\nabla_\mathbf{W}\ell$ of loss function $\ell$ with respect to the weights $\mathbf{W}$.
        \State $\mathbf{Q} \gets \mathbf{Q} + \text{sign}(\nabla_\mathbf{W}\ell)$
    \EndFor
    \State $\mathbf{U} \gets \mathbf{U} + \text{sign}(\mathbf{Q})$
    \If {${\text{max}(\lvert\mathbf{U}\rvert) = \rm{T}}$}
        \ForAll {$\lvert\mathbf{U}_{(i,j)}\rvert = \rm{T}$}
            \If{($\mathbf{U}_{(i,j)} = +\rm{T}$ and $\mathbf{W}_{(i,j)} = 0$) or \phantom{mm mm} \phantom{m} ($\mathbf{U}_{(i,j)} = -\rm{T}$ and $\mathbf{W}_{(i,j)} = 1$)}
                \State $\mathbf{W}_{(i,j)} \gets \mathbf{W}_{(i,j)} + 0$
            \Else
                \State $\mathbf{W}_{(i,j)} \gets \mathbf{W}_{(i,j)} - \frac{\mathbf{U}_{(i,j)}}{\rm{T}}$
            \EndIf
        \EndFor
    \State $\mathbf{U} \gets 0$
    \EndIf
\EndFor
\end{algorithmic}
\end{algorithm}

\subsection{Generalization to Non-Binary Discrete Levels}

We would like to highlight that these approaches could extend beyond binary values, and therefore address a  broader range of discrete optimization problems. Indeed, the quantization of the gradient as a technique for updating discrete weights is generic, depending primarily on the ability to provide relevant quantization thresholds to the algorithm. It is also important to note that the GQLA techniques do not interfere with the back-propagation process itself. As a result, scalar weights can be updated using standard optimization algorithms alongside the GQLA approach for other discrete weights. However, the applicability of these algorithms to non-binary problems is out of the scope of this paper.

% Quantization du gradient, accumulation des gradients quantifiés, prise de décision sur la base des valeurs accumulées. 3 paragraphes, un exemple dans le cas binaire à chaque fois.

\section{Training \& Evaluation Parameters}
\label{section:training_parameters}
The used hyper-parameters for each code rate are provided in Table~\ref{tab:HP_train}. The parameters $\alpha$, $N_{\rm{errors}}$, $\rm{T}$, and $\rm{D}$ (the 1's density in matrix $\mathbf{W}$ at initialization) greatly impact the capacity of the system to learn an efficient code. Only the right combination of these hyper-parameters provides the codes whose results are presented in Section~\ref{section:results}. Hence, a benchmark is necessary to test all possible hyper-parameters combinations in certain ranges. Hopefully, the training of a single code typically requires only a few minutes using an NVIDIA RTX A6000 GPU, allowing a quick coarse search of the right parameters, followed by a finer benchmark process. The coarse search is conducted with one training session per parameter combination for each code size to identify promising ranges. We then perform the finer analysis with five training sessions per combination within these ranges. Combinations are ranked based on BLER performance at maximum $\mathrm{E_b/N_0}$ and consistency across the five trainings. The final selection balances these criteria, optimizing both performance and reliability for each code size. From our experiments, reliable coarse search ranges are: $[1.2;5.0]$ with a step of $0.2$ for $\alpha$, $\{2,3,4,5,6\}$ for $\rm{N_{errors}}$, $\{10,20,30\}$ for $\rm{T}$, and $[0.15;0.45]$ with a step of $0.1$ for $\rm{D}$.

\begin{table}
    \centering
    \begin{tabular}{|c|c|c|c|c|}
        \hline
        $(n,k)$               & (32,16) & (64,32) & (64,16) & (128,64) \\
        \hline
        $\alpha$              &   2.5   &   2.7   &   1.6   &   2.8    \\
        \hline
        $N_{\rm{errors}}$     &    2    &    3    &    5    &    4     \\
        \hline
        $\rm{T}$              &    30   &    20   &    20   &    20    \\
        \hline
        $\rm{D}$              &   0.45  &   0.25  &   0.20  &   0.15   \\
        \hline
        Val. SNR [dB]         &    2    &    2    &    0    &    2     \\
        \hline
    \end{tabular}
    \caption{Hyper-parameters for each code rate.}
    \label{tab:HP_train}
\end{table}

The training is done on a maximum of 256 epochs with 100 steps per epochs. Thus, if the threshold of the Update Matrix is set to $\rm{T}=20$, then the weights can only be updated a maximum of 5 times per epoch.

The loss function is a binary cross entropy. Other loss functions presented in \cite{BLERvsBER} and focusing on the \ac{BLER} were tested, providing codes with the same performance.

Consistent with the findings reported in \cite{GNBP_Learning}, we observe superior performance in learned codes when training is conducted using 3 \ac{BP} iterations, even when the evaluation is performed with 5 iterations. An hypothesis is that this approach strikes a balance in the training process between efficient gradient propagation through the graph (facilitated by fewer iterations) and comprehensive consideration of message-passing steps in the loss function (requiring multiple iterations). To improve the gradient propagation, we replace the gradient of the inverse hyperbolic tangent (arctanh) operation within the \ac{BP} algorithm with a pass-through gradient. 

Finally, and as in \cite{GNBP_Learning}, the Agresti-Coull method \cite{BinomialConfidenceInterval} is used on the validation and evaluation \ac{BLER}. The validation step is stopped when there is a 95\% probability that the true value of the \ac{BLER} lies in-between $\pm$30\% of the estimated \ac{BLER}. The validation is done with 5 iterations and consider an \ac{AWGN} channel at the \ac{SNR} defined in Table~\ref{tab:HP_train}. Hence no scaling factor $\alpha$ is used and the true \ac{LLR}s are computed as input of the decoder. The chosen \ac{SNR} value optimizes the trade-off between achieving sufficiently low \ac{BLER} values for meaningful code performance comparison and meeting the Agresti-Coull confidence criteria within a computationally reasonable timeframe. The early stopping callback uses a patience of 10 epochs.

The evaluation step uses a complete system, where the generator matrix and the linear encoder are also implemented to evaluated random code-words. Based on the parity check matrix in standard form, the generator matrix is constructed as $\mathbf{G} = \left([\mathbf{W}^{(n-k),k}]^\mathrm{T}|\mathbf{I}^{k,k}\right)$. The encoder is described as a non-differentiable function computing the matrix multiplication between the data bits vector and the generator matrix in the Galois Field of cardinality 2. The \ac{BLER} metric is measured after 5 (or more if indicated) \ac{BP} iterations, with an \ac{AWGN} channel and no scaling factor $\alpha$ (true \ac{LLR}s) at the input of the decoder. The Agresti-Coull method is set to stop the evaluation when there is a 95\% probability that the true value of the \ac{BLER} lies in-between $\pm$10\% of the estimated \ac{BLER}.

\section{Simulations Results}
\label{section:results}

\subsection{Comparison with Random Codes}

To evaluate the consistency and efficiency of our code learning approach, we conducted five training sessions for each code size, considering batch sizes ($\rm{B}$) of 8 and 64. Table~\ref{tab:matrix_updates} summarizes the number of matrix updates for each training session across different code and batch sizes.

Our initial intention was to present a statistical comparison between learned and random codes, akin to Figure~\ref{fig:RvL}, following the methodology outlined in Section~\ref{section:code_comparison_methodology}. However, our findings reveal that every learned code consistently outperform the best randomly sampled one, underscoring the efficiency of our proposed code learning technique.

% The total number of codes processed during the five training sessions per code size is significantly lower than the 12,800 random samples. This disparity in sample sizes, coupled with the consistent superiority of learned codes, .

Moreover, we observe no significant performance difference between codes learned with batch sizes of 8 and 64. In the following, we present only the performance of the best learned codes, where the best code is determined based on its \ac{BLER} performance at the maximum $\mathrm{E_b/N_0}$. For GQLA techniques, the corresponding batch size is indicated in the legends.

%This finding suggests that our system can effectively learn efficient codes even with small batch sizes, potentially reducing the computational complexity of the approach and enabling the training of code with larger size. However, it's important to note that smaller batch sizes may increase the number of epochs required for convergence and could lead to less consistent code performance. For applications requiring enhanced performance consistency, we recommend increasing the batch size.

\begin{table}[htbp]
\centering
\begin{tabular}{|c|c|c|c|c|c|c|c|}
\hline
\multirow{2}{*}{Code Size} & \multirow{2}{*}{B} & \multicolumn{5}{c|}{Number of Updates per Session} & \multirow{2}{*}{Total} \\
\cline{3-7}
& & \#1 & \#2 & \#3 & \#4 & \#5 & \\
\hline
\hline
\multirow{2}{*}{(32,16)} & 8 & 64 & \cellcolor{gray!30}106 & 71 & 60 & 78 & 379 \\
\cline{2-8}
 & 64 & 74 & 67 & 107 & 115 & 121 & 484 \\
\hline
\hline
\multirow{2}{*}{(64,16)} & 8 & 102 & 133 & 140 & 145 & \cellcolor{gray!30}122 & 642 \\
\cline{2-8}
 & 64 & 86 & 117 & 79 & 134 & \cellcolor{gray!30}169 & 585 \\
\hline
\hline
\multirow{2}{*}{(64,32)} & 8 & 197 & 199 & 286 & \cellcolor{gray!30}207 & 182 & 1071 \\
\cline{2-8}
 & 64 & 216 & 147 & 158 & \cellcolor{gray!30}176 & 226 & 923 \\
\hline
\hline
\multirow{2}{*}{(128,64)} & 8 & 239 & 267 & 367 & 329 & \cellcolor{gray!30}314 & 1516 \\
\cline{2-8}
 & 64 & 289 & 285 & 306 & \cellcolor{gray!30}289 & 268 & 1437 \\
\hline
\end{tabular}
\captionsetup{justification=centering}
\caption{Number of matrix updates for each training session, per code and batch size. Gray cells are the codes whose results are presented in the following sub-sections.}
\label{tab:matrix_updates}
\end{table}

\subsection{(32,16) Codes - Comparison of MB-GQLA and S-GQLA Approaches}

    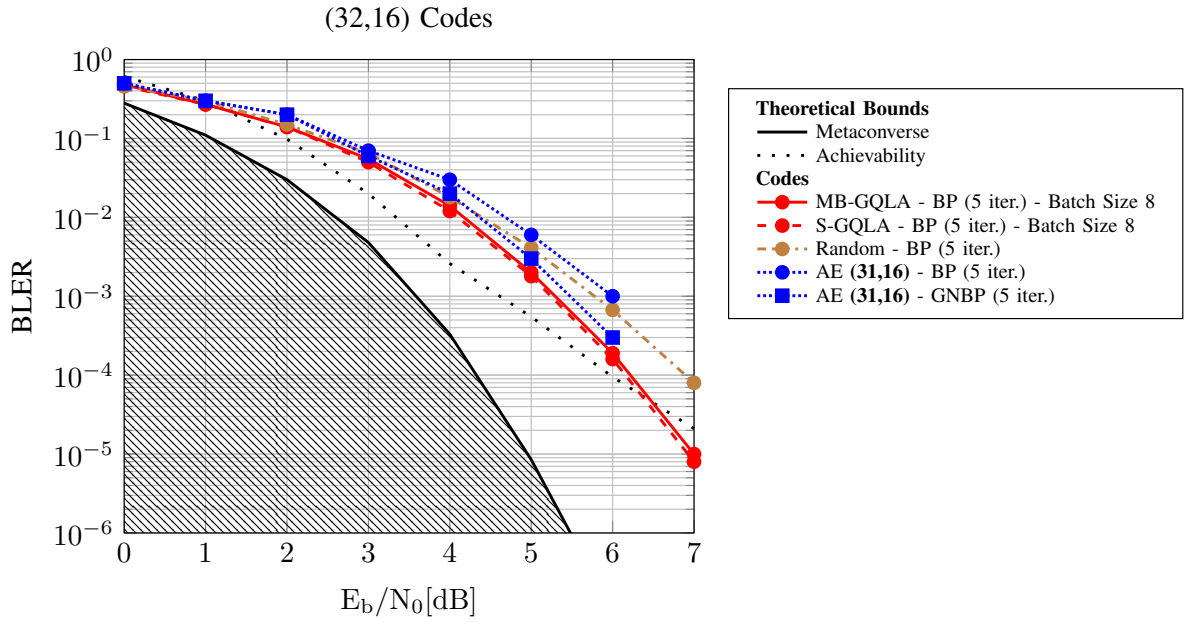
\begin{figure*}[htbp]
        \centering
        \scalebox{1.1}{\begin{tikzpicture}[scale=1]
    %Manually reduce line spacing to single space no matter double or single column document (draft mode)
    \linespread{1}

    \begin{axis}[
        title={(32,16) Codes},
        title style = {align=center},
        xmin=0, 
        xmax=7, 
        xminorgrids=true,
        xmajorgrids=true,
        %xmode=log,
        xlabel = \(\mathrm{E_b/N_0 [dB]}\), 
        ymin=1e-6, 
        ymax=1, 
        yminorgrids=true,
        ymajorgrids=true,
        ymode=log,
        ylabel = BLER,
        legend cell align=left,
    ]  
        %%%%%%%%%%%%%%%%%%%%%%%%%%%%%%%% DATASETS %%%%%%%%%%%%%%%%%%%%%%%%%%%%%%%%%%%%%%%%%%%%%%%%%%%%%%%
        %XP
        %Eb/No      ZC (5 iter) /   ZC Jacobian (5 iter)    /   Achievability   /   Metaconverse    /   Random
        \pgfplotstableread{
            0       4.8e-1          4.6e-1                      6.1e-1              2.8e-1              4.6e-1
            1       2.7e-1          2.7e-1                      3.0e-1              1.1e-1              3.0e-1
            2       1.4e-1          1.4e-1                      9.8e-2              3.0e-2              1.5e-1
            3       5.5e-2          5.0e-2                      2.0e-2              4.8e-3              6.5e-2
            4       1.4e-2          1.2e-2                      2.6e-3              3.3e-4              1.8e-2
            5       2.0e-3          1.8e-3                      5.5e-4              8.5e-6              4.1e-3
            6       1.9e-4          1.6e-4                      9.5e-5              1.0e-7              6.7e-4
            7       1.0e-5          8.0e-6                      2.1e-5              nan                 8e-5
        }\datasetXP

        %Previous /!\ (31,16) not (32,16) /!\
        %Eb/No      AE GNBP (5 iter)    /   AE BP (5 iter)   / Choukroun et al. 
        \pgfplotstableread{
            0       5.0e-1                  5.0e-1      nan                                 
            1       3.0e-1                  3.0e-1      nan                                   
            2       2.0e-1                  2.0e-1      nan                               
            3       6.0e-2                  7.0e-2      5e-2                                 
            4       2.0e-2                  3.0e-2      1.6e-2                                 
            5       3.0e-3                  6.0e-3      3e-3                                 
            6       3.0e-4                  1.0e-3      4.5e-4                                    
            7       nan                     nan         5e-5                             
        }\datasetPrevious

        %%%%%%%%%%%%%%%%%%%%%%%%%%%%%%%%%%%%%%%%%%%%%%%%%%%%%%%%%%%%%%%% PLOTS %%%%%%%%%%%%%%%%%%%%%%%%%%%%%%%%%%%%%%%%%%%%%

        %BOUNDS
        \addplot[color=black,line width=1pt,solid, mark size=2pt, mark=empty, mark options={solid}, name path=MCV] table [x=0, y=4] \datasetXP;
        \label{plot:3216MCV}
        
        \addplot[color=black,line width=1pt,loosely dotted, mark size=2pt, mark=empty, mark options={solid}] table [x=0, y=3] \datasetXP;
        \label{plot:3216Achievability}

        %Hatches bellow MCV
        \addplot+[draw=none,name path=B, domain=0:6, mark=none] {1e-6};     
        \addplot+[gray, pattern=north west lines] fill between[of=MCV and B];

        % %Zero Coder
        \addplot[color=red,line width=1pt,solid, mark size=2pt, mark=*, mark options={solid}] table [x=0, y=1] \datasetXP;
        \label{plot:3216ZC}

        \addplot[color=red,line width=1pt,dashed, mark size=2pt, mark=*, mark options={solid}] table [x=0, y=2] \datasetXP;
        \label{plot:3216ZCJC}
        
        \addplot[color=brown,line width=1pt,dashdotted, mark size=2pt, mark=*, mark options={solid}] table [x=0, y=5] \datasetXP;
        \label{plot:3216Random}

        % %AE
        \addplot[color=blue,line width=1pt,densely dotted, mark size=2pt, mark=square*, mark options={solid}] table [x=0, y=1] \datasetPrevious;
        \label{plot:3116AEGNBP}
        
        \addplot[color=blue,line width=1pt,densely dotted, mark size=2pt, mark=*, mark options={solid}] table [x=0, y=2] \datasetPrevious;
        \label{plot:3116AEBP}

        % % Choukroun et al
        % \addplot[color=Orange,line width=1pt,dashed, mark size=2pt, mark=star, mark options={solid}] table [x=0, y=3] \datasetPrevious;
        % \label{plot:3216Choukroun}

    \end{axis}
    
    %Custom Legend %at (3,-1.25)
    \node [draw,fill=white,anchor=north,font=\scriptsize] at (10,5.35) {
        \begin{tabular}{l}
        \textbf{Theoretical Bounds}\\
        \ref{plot:3216MCV} Metaconverse\\
        \ref{plot:3216Achievability} Achievability\\

        \textbf{Codes}\\
        \ref{plot:3216ZC} MB-GQLA - BP (5 iter.) - Batch Size 8\\ 
        \ref{plot:3216ZCJC} S-GQLA - BP (5 iter.) - Batch Size 8\\ 
        \ref{plot:3216Random} Random - BP (5 iter.)\\     
        \ref{plot:3116AEBP} AE \textbf{(31,16)} - BP (5 iter.)\\ 
        \ref{plot:3116AEGNBP} AE \textbf{(31,16)} - GNBP (5 iter.)\\  
        % \ref{plot:3216Choukroun} \textcolor{red}{Choukroun et al. \cite{Choukroun2024b} - BP (5 iter.)}\\    

        \end{tabular}
    };
\end{tikzpicture}}
        \captionsetup{justification=centering}
        \caption{Codes (32,16) with S-/MB-GQLA quantization techniques. Comparison with the best random code and the auto-encoder results from our previous work \cite{GNBP_Learning} on (31,16) codes.}%and results from \textcolor{red}{Choukroun et al. \cite{Choukroun2024b}}.
        \label{fig:3216}
    \end{figure*}
    
    We conducted a comparative study to evaluate the performance of our proposed approaches, MB-GQLA and S-GQLA with the Update Matrix, for a (32,16) code size. Figure~\ref{fig:3216} presents the BLER results, comparing these approaches against the best random code and the previously proposed Auto-Encoder (AE) model \cite{GNBP_Learning}. We include two finite block length limits from \cite{Polyanskiy}: the Metaconverse (a lower bound) and the Achievability limit (an upper bound). These bounds indicate the theoretical attainable performance range for the specified code size.
    
    Our findings indicate that both MB-GQLA and S-GQLA approaches yield comparable performance levels. While S-GQLA potentially offers finer exploitation of individual sample gradient terms, its current implementation introduces additional complexity without discernible performance gains. Consequently, we focus solely on the simpler MB-GQLA approach for the remainder of this paper.
    
    Our proposed approaches demonstrate superior performance compared to the AE BP model from \cite{GNBP_Learning}. The performance gap narrows when comparing with the AE GNBP model, which incorporates trainable weights in the decoder graph.
    
    It's important to note several caveats in this comparison, that advantage the AE GNBP approach:
    \begin{itemize}
        \item The code sizes differ slightly: (32,16) in our work versus (31,16) in \cite{GNBP_Learning}, due to practical considerations.
        \item The total number of matrix updates in \cite{GNBP_Learning} exceeds that in our current work.  
        \item The AE GNBP model's performance may be partially attributed to its trainable decoder, which our approach does not incorporate.
    \end{itemize}

\subsection{(64,16) Codes - Comparison with DSF}

   \begin{figure*}[htbp]
        \centering
        \scalebox{1.1}{\begin{tikzpicture}[scale=1]
    %Manually reduce line spacing to single space no matter double or single column document (draft mode)
    \linespread{1}

    \begin{axis}[
        title={(64,16) Codes},
        title style = {align=center},
        xmin=0, 
        xmax=7, 
        xminorgrids=true,
        xmajorgrids=true,
        %xmode=log,
        xlabel = \(\mathrm{E_b/N_0 [dB]}\), 
        ymin=1e-6, 
        ymax=1, 
        yminorgrids=true,
        ymajorgrids=true,
        ymode=log,
        ylabel = BLER,
        legend cell align=left,
    ]  
        %%%%%%%%%%%%%%%%%%%%%%%%%%%%%%%% DATASETS %%%%%%%%%%%%%%%%%%%%%%%%%%%%%%%%%%%%%%%%%%%%%%%%%%%%%%%
        %XP
        %Eb/No      ZC (5 iter) /   Achievability   /   Metaconverse    /   Random     /  ZC (5 ite + BS 64) 
        \pgfplotstableread{
            0       3.9e-1          3.0e-1              1.4e-1              3.8e-1          4.0e-1
            1       2.2e-1          9.6e-2              4.0e-2              2.2e-1          2.2e-1
            2       9.2e-2          1.9e-2              7.0e-3              1.0e-1          9.3e-2
            3       3.2e-2          2.1e-3              5.5e-4              4.0e-2          2.8e-2
            4       6.7e-3          1.2e-4              1.6e-5              1.3e-2          6.6e-3
            5       9.8e-4          5.5e-6              1e-7                2.3e-3          8.7e-4
            6       1.0e-4          nan                 nan                 4.0e-4          8.9e-5
            7       6.5e-6          nan                 nan                 4.0e-5          5.9e-6
        }\datasetXP

        %BS64:1496 BS1280:1097 updates
        %Eb/No      DSF-Id-64 (5 iter)    /   DSF-Id-1280 (5 iter)  /   Irregular LDPC - BP (5iter) /   Irregular LDPC - GNBP (5iter)        
        \pgfplotstableread{
            0       4.0e-1                  3.9e-1              6.0e-1                          6.0e-1    
            1       2.3e-1                  2.2e-1              4.0e-1                          4.0e-1      
            2       8.4e-2                  1.1e-1              3.0e-1                          3.0e-1  
            3       3.2e-2                  4.4e-2              1.0e-1                          8.0e-2    
            4       6.3e-3                  7.3e-3              3.0e-2                          2.0e-2    
            5       9.2e-4                  1.1e-3              4.0e-3                          3.0e-3    
            6       1.2e-4                  1.2e-4              3.0e-4                          2.0e-4       
            7       7.7e-6                  6.8e-6                 nan                             nan
        }\datasetPrevious

        %%%%%%%%%%%%%%%%%%%%%%%%%%%%%%%%%%%%%%%%%%%%%%%%%%%%%%%%%%%%%%%% PLOTS %%%%%%%%%%%%%%%%%%%%%%%%%%%%%%%%%%%%%%%%%%%%%

        %BOUNDS
        \addplot[color=black,line width=1pt,solid, mark size=2pt, mark=empty, mark options={solid}, name path=MCV] table [x=0, y=3] \datasetXP;
        \label{plot:6416MCV}
        
        \addplot[color=black,line width=1pt,loosely dotted, mark size=2pt, mark=empty, mark options={solid}] table [x=0, y=2] \datasetXP;
        \label{plot:6416Achievability}

        %Hatches bellow MCV
        \addplot+[draw=none,name path=B, domain=0:6, mark=none] {1e-6};     
        \addplot+[gray, pattern=north west lines] fill between[of=MCV and B];

        % %Zero Coder
        \addplot[color=red,line width=1pt,solid, mark size=2pt, mark=*, mark options={solid}] table [x=0, y=1] \datasetXP;
        \label{plot:6416ZC}
        \addplot[color=red,line width=1pt,solid, mark size=2pt, mark=triangle*, mark options={solid}] table [x=0, y=5] \datasetXP;
        \label{plot:6416ZC-64}

        \addplot[color=brown,line width=1pt,dashdotted, mark size=2pt, mark=*, mark options={solid}] table [x=0, y=4] \datasetXP;
        \label{plot:6416Random}

        % %AE
        \addplot[color=blue,line width=1pt,densely dotted, mark size=2pt, mark=square*, mark options={solid}] table [x=0, y=2] \datasetPrevious;
        \label{plot:6416DSF-1280}
        
        \addplot[color=blue,line width=1pt,densely dotted, mark size=2pt, mark=*, mark options={solid}] table [x=0, y=1] \datasetPrevious;
        \label{plot:6416DSF-64}

        % LDPC
        % \addplot[color=ForestGreen,line width=1pt,dashed, mark size=2pt, mark=*, mark options={solid}] table [x=0, y=3] \datasetPrevious;
        % \label{plot:6418LDPCBP}

        % \addplot[color=ForestGreen,line width=1pt,dashed, mark size=2pt, mark=square*, mark options={solid}] table [x=0, y=4] \datasetPrevious;
        % \label{plot:6418LDPCGNBP}

    \end{axis}
    
    %Custom Legend %at (3,-1.25)
    \node [draw,fill=white,anchor=north,font=\scriptsize] at (10,5.35) {
        \begin{tabular}{l}
        \textbf{Theoretical Bounds}\\
        \ref{plot:6416MCV} Metaconverse\\
        \ref{plot:6416Achievability} Achievability\\

        \textbf{Codes}\\
        \ref{plot:6416ZC} MB-GQLA - BP (5 iter.) - Batch Size 8\\ 
        \ref{plot:6416ZC-64} MB-GQLA - BP (5 iter.) - Batch Size 64\\ 
        \ref{plot:6416Random} Random - BP (5 iter.)\\     
        \ref{plot:6416DSF-64} DSF - Batch Size 64 (5 iter.)\\ 
        \ref{plot:6416DSF-1280} DSF - Batch Size 1280 (5 iter.)\\  
        %\ref{plot:6418LDPCBP} Irregular LDPC \textbf{(64,18)} - BP (5 iter.)\\ 
        %\ref{plot:6418LDPCGNBP} Irregular LDPC \textbf{(64,18)} - GNBP (5 iter.)\\     

        \end{tabular}
    };
\end{tikzpicture}}
        \caption{Codes (64,16), comparison with the best random code and the DSF approach.}
        \label{fig:6416}
    \end{figure*}
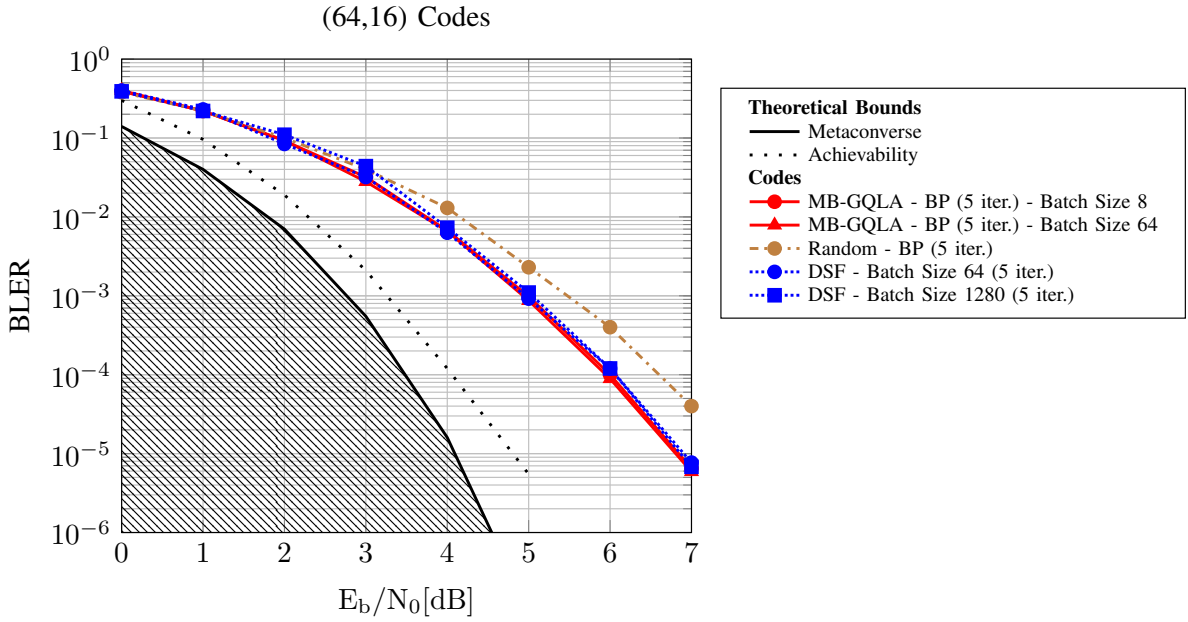

    This section examines the performance of a more conventional training approach using real-valued weights associated with the \ac{DSF} defined in Equation~\ref{eq:dsf}. We employ the classic mini-batch gradient descent algorithm with a scaling factor set to $1$, hence not including gradient momentum from previously evaluated codes over the course of training.

    As with the GQLA approaches, hyper-parameters configuration plays a crucial role in the final performance of the learned code. We adjusted the values of $\alpha$, $N_{\rm{errors}}$, and $\rm{D}$, as well as the weight initialization values. Since the weights are no longer binary-valued, they can be initialized anywhere in the range $[-\infty, +\infty]$. Inspired by the results on binary weights, we opted for a symmetric initialization where each weight is set to either $+\rm{V}$ or $-\rm{V}$ with a probability depending on $\rm{D}$. After a benchmarking process similar to that used for GQLA, we arrived at the following configuration: $\alpha=1.4$, $N_{\rm{errors}}=4$, $\rm{D}=0.15$, and $\rm{V}=\pm1e^{-3}$.
    
    Figure~\ref{fig:6416} presents the performance results of the DSF and MB-GQLA approaches. Both methods achieve equivalent performance, outperforming random search. For each approach, we selected the best of five learned codes.

    As explained in Section~\ref{section:gradient_quantificatin_mechanisms}, our binary-valued system treats weights as two real values (0 and 1) with a continuous identity extension over [0,1], enabling differentiable graph and gradient computation. In the case of the DSF approach, a pass-through gradient approximation is used, effectively treating the step function as an identity function during back-propagation. Consequently, the computed and back-propagated gradients are mathematically equal in both approaches. The key difference lies in the actual update values applied to the weights by the different optimizer algorithms.

    The fact that both approaches achieve similar performance levels, despite different update mechanisms, suggests the central role of the system model and design choices (error channel, scaling factor $\alpha$, Agresti-Coull based validation, etc.) in the learning process. Moreover, the decisive impact of the initial weight value $\mathrm{V}$ on the DSF approach's ability to learn performing codes underscores the importance of numerical balance in the system. This observation aligns with our findings regarding the impact of $\alpha$ on GQLA performance, indicating that careful calibration of gradient magnitudes throughout training is crucial for effective learning with both binary and real-valued weights. The role of $\mathrm{V}$ in the DSF approach can also be seen as analogous to the threshold $\mathrm{T}$ in the matrix update approach, in that larger $\mathrm{V}$ values require more incremental updates for weights to cross the step function's threshold.

    To evaluate the impact of gradient's estimation accuracy, we tested the DSF approach with batch sizes of 64 and 1280 samples. Intriguingly, both consistently provided codes with equivalent performances. To further investigate, we employ the Agresti-Coull method during training to dynamically determine the sufficient number of batches for a reliable BLER estimation. We hypothesize that a reliable BLER estimate would correspond to a reliable gradient estimate. For the MB-GQLA, we adapt our approach by implementing a dynamic threshold for the Update Matrix: once the Agresti-Coull method is triggered, we only update weights whose accumulators have reached the current maximum value in the Update Matrix. For the DSF, we average the gradient over all involved batches, thus effectively employing a larger batch size. In this configuration, both DSF and GQLA approaches successfully learn (64,16) codes, achieving performance equivalent to those presented earlier. These findings may suggest that the best known configuration of hyper-parameters results in gradient values sufficiently accurate for training convergence.

\subsection{(64,32) Codes - Comparison with irregular LDPC Codes}
    
    \begin{figure*}[htbp]%\begin{figure}[hb]
        \centering
        \scalebox{1.1}{\begin{tikzpicture}[scale=1]
    %Manually reduce line spacing to single space no matter double or single column document (draft mode)
    \linespread{1}

    \begin{axis}[
        title={(64,32) Codes},
        title style = {align=center},
        xmin=0, 
        xmax=7, 
        xminorgrids=true,
        xmajorgrids=true,
        %xmode=log,
        xlabel = \(\mathrm{E_b/N_0 [dB]}\), 
        ymin=1e-7, 
        ymax=1, 
        yminorgrids=true,
        ymajorgrids=true,
        ymode=log,
        ylabel = BLER,
        legend cell align=left,
    ]  
        %%%%%%%%%%%%%%%%%%%%%%%%%%%%%%%% DATASETS %%%%%%%%%%%%%%%%%%%%%%%%%%%%%%%%%%%%%%%%%%%%%%%%%%%%%%%
        %XP
        %Eb/No      ZC (5 iter) /   Achievability   /   Metaconverse    /   Random      /   ZC (5 ite. + BS64)
        \pgfplotstableread{
            0       5.9e-1          6e-1                3.2e-1              6e-1                6.3e-1
            1       3.9e-1          2.2e-1              1e-1                4e-1                3.8e-1
            2       1.7e-1          4e-2                1.6e-2              2e-1                1.6e-1
            3       4.4e-2          2.8e-3              8e-4                6e-2                4.0e-2
            4       7.1e-3          8e-5                1e-5                1.5e-2              6.0e-3
            5       7e-4            2.2e-6              5e-8                2.4e-3              4.8e-4
            6       4.1e-5          1e-7                nan                 2.3e-4              2.3e-5
            7       1.4e-6          nan                 nan                 1.5e-5              6.2e-7
        }\datasetXP

        %Previous
        %Eb/No      AE GNBP (5 iter)    /   AE BP (5 iter)  /   Irregular LDPC - BP (5iter) /   Irregular LDPC - GNBP (5iter)    / Choukroun et al.
        \pgfplotstableread{
            0       8e-1                    8e-1                8e-1    8e-1    nan                        
            1       6e-1                    6e-1                5e-1    5e-1    nan                           
            2       3e-1                    3.5e-1              3e-1    3e-1    nan 
            
            3       1.1e-1                  1.7e-1              8e-2    8e-2    5.5e-2                           
            4       3e-2                    5.5e-2              2e-2    2e-2    9e-3                          
            5       4e-3                    1.5e-2              2.3e-3  2e-3    1.2e-3                           
            6       5e-4                    4e-3                2.3e-4  2e-4    9e-5                           
            7       nan                     nan                 nan     nan     6.5e-6                      
        }\datasetPrevious

        %%%%%%%%%%%%%%%%%%%%%%%%%%%%%%%%%%%%%%%%%%%%%%%%%%%%%%%%%%%%%%%% PLOTS %%%%%%%%%%%%%%%%%%%%%%%%%%%%%%%%%%%%%%%%%%%%%

        %BOUNDS
        \addplot[color=black,line width=1pt,solid, mark size=2pt, mark=empty, mark options={solid}, name path=MCV] table [x=0, y=3] \datasetXP;
        \label{plot:6432MCV}
        
        \addplot[color=black,line width=1pt,loosely dotted, mark size=2pt, mark=empty, mark options={solid}] table [x=0, y=2] \datasetXP;
        \label{plot:6432Achievability}

        %Hatches bellow MCV
        \addplot+[draw=none,name path=B, domain=0:6, mark=none] {1e-7};     
        \addplot+[gray, pattern=north west lines] fill between[of=MCV and B];

        % %Zero Coder
        \addplot[color=red,line width=1pt,solid, mark size=2pt, mark=*, mark options={solid}] table [x=0, y=1] \datasetXP;
        \label{plot:6432ZC}

        \addplot[color=red,line width=1pt,solid, mark size=2pt, mark=triangle*, mark options={solid}] table [x=0, y=5] \datasetXP;
        \label{plot:6432ZC-64}
        
        \addplot[color=brown,line width=1pt,dashdotted, mark size=2pt, mark=*, mark options={solid}] table [x=0, y=4] \datasetXP;
        \label{plot:6432Random}

        % %AE
        \addplot[color=blue,line width=1pt,densely dotted, mark size=2pt, mark=square*, mark options={solid}] table [x=0, y=1] \datasetPrevious;
        \label{plot:6432AEGNBP}
        
        \addplot[color=blue,line width=1pt,densely dotted, mark size=2pt, mark=*, mark options={solid}] table [x=0, y=2] \datasetPrevious;
        \label{plot:6432AEBP}

        % LDPC
        \addplot[color=ForestGreen,line width=1pt,dashed, mark size=2pt, mark=*, mark options={solid}] table [x=0, y=3] \datasetPrevious;
        \label{plot:6432LDPCBP}

        \addplot[color=ForestGreen,line width=1pt,dashed, mark size=2pt, mark=square*, mark options={solid}] table [x=0, y=4] \datasetPrevious;
        \label{plot:6432LDPCGNBP}

        % Choukroun et al
        \addplot[color=Orange,line width=1pt,dashed, mark size=2pt, mark=star, mark options={solid}] table [x=0, y=5] \datasetPrevious;
        \label{plot:6432Choukroun}

    \end{axis}
    
    %Custom Legend %at (3,-1.25)
    \node [draw,fill=white,anchor=north,font=\scriptsize] at (10,5.35) {
        \begin{tabular}{l}
        \textbf{Theoretical Bounds}\\
        \ref{plot:6432MCV} Metaconverse\\
        \ref{plot:6432Achievability} Achievability\\

        \textbf{Codes}\\
        \ref{plot:6432ZC} MB-GQLA - BP (5 iter.) - Batch Size 8\\ 
        \ref{plot:6432ZC-64} MB-GQLA - BP (5 iter.) - Batch Size 64\\ 
        \ref{plot:6432Random} Random - BP (5 iter.)\\     
        \ref{plot:6432AEBP} AE - BP (5 iter.)\\ 
        \ref{plot:6432AEGNBP} AE - GNBP (5 iter.)\\     
        \ref{plot:6432LDPCBP} Irregular LDPC - BP (5 iter.)\\ 
        \ref{plot:6432LDPCGNBP} Irregular LDPC - GNBP (5 iter.)\\   
        \ref{plot:6432Choukroun} Choukroun et al. \cite{Choukroun2024b} - BP (5 iter.)\\    

        \end{tabular}
    };
\end{tikzpicture}}
        \caption{Codes (64,32), comparison with the best random code, results from our previous work \cite{GNBP_Learning}, Choukroun et al. \cite{Choukroun2024b} and irregular LDPC.}
        \label{fig:6432}
    \end{figure*}
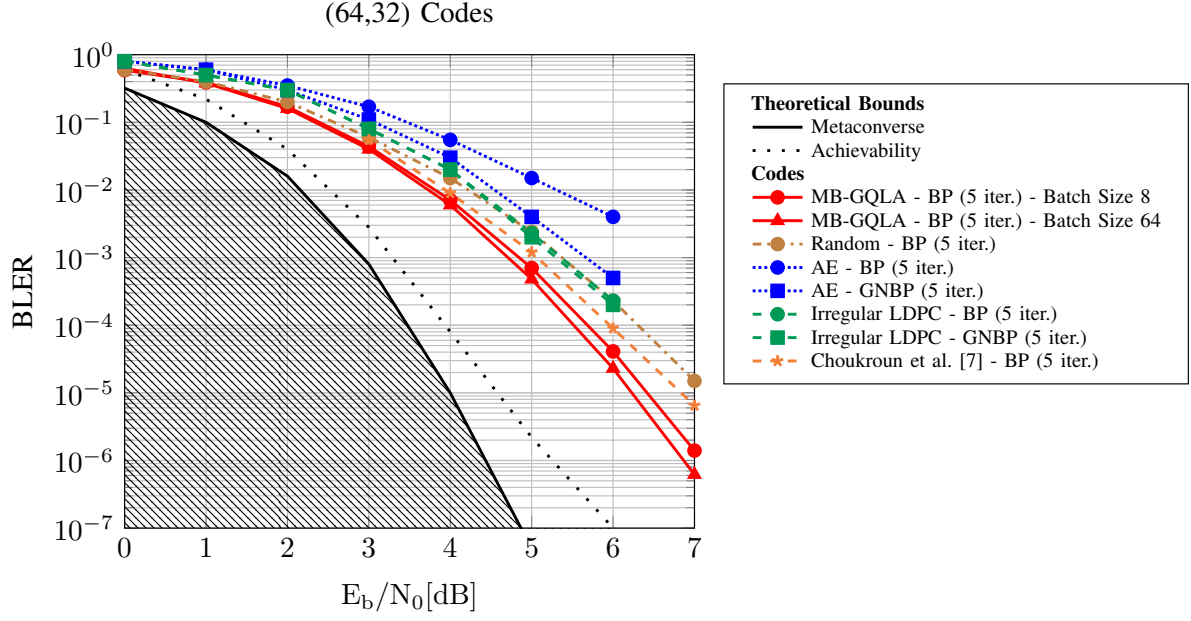%\end{figure}
    
    This section presents a comprehensive performance evaluation of our proposed approach for code size (64,32). We benchmark our results against previous work \cite{GNBP_Learning}, state-of-the-art irregular LDPC codes constructed using Progressive Edge Growth (PEG) methods \cite{peg}, and random code search results. The comparison encompasses codes decoded using both a standard BP decoder and a trainable GNBP decoder. Additionally, we include results from recent work by Choukroun et al. \cite{Choukroun2024b}, who also seek to learn short codes optimized for BP decoding. Their results provide a valuable reference point for evaluating ML-based code design approaches.

    Figure~\ref{fig:6432} illustrates that our proposed model consistently outperforms all other codes tested in this study across all decoding strategies. The improvement over both learned (AE and \cite{Choukroun2024b}) and expert-crafted (PEG) codes highlights the potential of our approach in designing state-of-the-art codes for small code sizes when utilizing BP decoding.

\subsection{(128,64) Codes - Scalability and Comparison with SotA Codes}
    
    \begin{figure*}[htbp]%\begin{figure}[hb]
        \centering
        \scalebox{1.1}{\begin{tikzpicture}[scale=1]
    %Manually reduce line spacing to single space no matter double or single column document (draft mode)
    \linespread{1}

    \begin{axis}[
        title={(128,64) Codes},
        title style = {align=center},
        xmin=1, 
        xmax=6, 
        xminorgrids=true,
        xmajorgrids=true,
        %xmode=log,
        xlabel = \(\mathrm{E_b/N_0 [dB]}\), 
        ymin=1e-4, 
        ymax=1, 
        yminorgrids=true,
        ymajorgrids=true,
        ymode=log,
        ylabel = BLER,
        legend cell align=left,
    ]  
    %%%%%%%%%%%%%%%%%%%%%%%%%%%%%%%% DATASETS %%%%%%%%%%%%%%%%%%%%%%%%%%%%%%%%%%%%%%%%%%%%%%%%%%%%%%%
        % Survey paper:
        %   Eb/No / SPB /       NA /    RCUB /      Ext. BCH OSD /  CCSD LDPC / AR3A LDPC / Polar LD +  CRC /   Polar SC /  TB CC m=8 / TB CC m=14 / Choukroun et al.
        \pgfplotstableread{
            1       8e-2        1e-1    7e-1        1.6e-1          nan         5e-1        1.8e-1              nan         1.5e-1      1.3e-1      nan      
            1.5     1.6e-2      3.1e-2  3e-1        5e-2            nan         2.5e-1      6e-2                nan         6e-2        3.2e-2      nan
            2       2.5e-3      7e-3    1e-1        1e-2            4e-1        1.2e-1      1.8e-2              2e-1        1.8e-2      7e-3        nan
            2.5     2e-4        8.5e-4    2.2e-2      1.1e-3        nan         4e-2        3e-3                nan         3.9e-3      9e-4        nan
            3       1e-5        6e-5    2.4e-3      1.3e-4          6e-2        1e-2        4.5e-4              2.2e-2        8e-4        9e-5      3.7e-2
            3.5     3e-7        2e-6    1.5e-4      8e-6            nan         2.2e-3      4.7e-5              nan         1.5e-4      5.5e-6      1e-2
            4       nan         nan     4.5e-6      5e-7            2.5e-3      3.3e-4      3.7e-6              2.4e-3      2.8e-5      nan         3e-3
            4.5     nan         nan     nan         nan             nan         5.2e-5      5e-7                nan         3.1e-6      nan         7e-4
            5       nan         nan     nan         nan             4e-5        5e-6        nan                 1.9e-4      5e-6        nan         1.7e-4
            5.5     nan         nan     nan         nan             nan         1e-6        nan                 nan         nan         nan         3.5e-5
            6       nan         nan     nan         nan             1e-7        nan         nan                 7e-6        nan         nan         8e-6
        }\dataset

        %XP
        %Eb/No      ZC (5 iter) /   AE GNBP (BEST 5 iter)   /   CCSDS (5iter)   /   ZC (200 iter)   / ZC (5 ite. + BS 64)   / ZC (200 ite. + BS 64)
        \pgfplotstableread{
            1       5.4e-1          9e-1                        9e-1                3.7e-1              5.4e-1                      3.6e-1
            2       2e-1            5e-1                        5.5e-1              9.7e-2              2.0e-1                      9.2e-2
            3       4e-2            3e-1                        2.1e-1              1.0e-2              3.4e-2                      9.8e-3
            4       4.3e-3          5e-2                        2.5e-2              9.7e-4              3.1e-3                      9.2e-4
            5       2.1e-4          5.2e-3                      9e-4                8.9e-5              1.4e-4                      7.9e-5
            6       nan             8e-5                        2e-5                nan                 nan                         nan
            
        }\datasetXP

        %%%%%%%%%%%%%%%%%%%%%%%%%%%%%%%%%%%%%%%%%%%%%%%%%%%%%%%%%%%%%%%% PLOTS %%%%%%%%%%%%%%%%%%%%%%%%%%%%%%%%%%%%%%%%%%%%%

        %BOUNDS
        \addplot[color=black,line width=1pt,solid, mark size=2pt, mark=empty, mark options={solid}, name path=SPB] table [x=0, y=1] \dataset;
        \label{plot:12864SPB}
        
        % \addplot[color=black,line width=1pt,loosely dashed, mark size=2pt, mark=empty, mark options={solid}] table [x=0, y=2] \dataset;
        % \label{plot:12864NA}
        
        \addplot[color=black,line width=1pt,loosely dotted, mark size=2pt, mark=empty, mark options={solid}] table [x=0, y=3] \dataset;
        \label{plot:12864RCUB}
        
        %Hatches bellow SPB
        \addplot+[draw=none,name path=B, domain=0:6, mark=none] {1e-5};     
        \addplot+[gray, pattern=north west lines] fill between[of=SPB and B];

        % %LDPC
                
        \addplot[color=blue,line width=1pt,densely dashed, mark size=2pt, mark=*, mark options={solid, fill=white}] table [x=0, y=6] \dataset;
        \label{plot:12864AR3ALDPC} % orange
        
        \addplot[color=ForestGreen,line width=1pt,dashed, mark size=2pt, mark=*, mark options={solid}] table [x=0, y=5] \dataset;
        \label{plot:12864CCSDLDPC} % ForestGreen,fill=white

        % \addplot[color=ForestGreen,line width=1pt,dashed, mark size=2pt, mark=*, mark options={solid}] table [x=0, y=3] \datasetXP;
        % \label{plot:12864CCSDSLDPCNSYSBP5iter}

        % Polar
        % \addplot[color=Plum,line width=1pt,dashdotdotted, mark size=2pt, mark=triangle*, mark options={solid}] table [x=0, y=8] \dataset;
        % \label{plot:12864PolarSC}

        %Zero Coder
        \addplot[color=red,line width=1pt,solid, mark size=2pt, mark=*, mark options={solid}] table [x=0, y=1] \datasetXP;
        \label{plot:12864ZC}

        \addplot[color=red,line width=1pt,solid, mark size=2pt, mark=triangle*, mark options={solid}] table [x=0, y=4] \datasetXP;
        \label{plot:12864ZC200iter} % red ,fill=white

        \addplot[color=Plum,line width=1pt,dashdotdotted, mark size=2pt, mark=*, mark options={solid}] table [x=0, y=5] \datasetXP;
        \label{plot:12864ZC-64}
        \addplot[color=Plum,line width=1pt,dashdotdotted, mark size=2pt, mark=triangle*, mark options={solid}] table [x=0, y=6] \datasetXP;
        \label{plot:12864ZC200iter-64}

        % AE 
        \addplot[color=blue,line width=1pt,densely dotted, mark size=2pt, mark=square*, mark options={solid}] table [x=0, y=2] \datasetXP;
        \label{plot:12864AEGNBP3iter5iter}

        % Choukroun et al
        \addplot[color=Orange,line width=1pt,dashed, mark size=2pt, mark=star, mark options={solid}] table [x=0, y=11] \dataset;
        \label{plot:12864Choukroun}

    \end{axis}
    
    %Custom Legend %at (3,-1.25)
    \node [draw,fill=white,anchor=north,font=\scriptsize] at (10,5.35) {
        \begin{tabular}{l}
        \textbf{Theoretical Bounds}  \\%\cite{CodeDesignForShortBlocksSurvey}\\
        \ref{plot:12864SPB} Sphere Packing\\
        %\ref{plot:12864NA}  Normal Approximation\\ 
        \ref{plot:12864RCUB} Achievability\\

        \textbf{Codes}\\
        \ref{plot:12864AEGNBP3iter5iter} AE - GNBP (5 iter.)\\ 
        \ref{plot:12864ZC} MB-GQLA - BP (5 iter.) - Batch Size 8\\
        \ref{plot:12864ZC200iter} MB-GQLA - BP (200 iter.) - Batch Size 8\\ 
        \ref{plot:12864ZC-64} MB-GQLA - BP (5 iter.) - Batch Size 64\\
        \ref{plot:12864ZC200iter-64} MB-GQLA - BP (200 iter.) - Batch Size 64\\ 
        %\ref{plot:12864CCSDSLDPCNSYSBP5iter} LDPC CCSDS - BP (5 iter.)\\      
        \ref{plot:12864CCSDLDPC} LDPC CCSDS - BP (200 iter.) \\%\cite{CodeDesignForShortBlocksSurvey} \\
        \ref{plot:12864AR3ALDPC} LDPC AR3A - BP (200 iter.) \\%\cite{CodeDesignForShortBlocksSurvey}\\
        %\ref{plot:12864PolarSC}  Polar Code - SC Decoding \cite{CodeDesignForShortBlocksSurvey} \\   
        \ref{plot:12864Choukroun} Choukroun et al. \cite{Choukroun2024b} - BP (5 iter.)\\     

        \end{tabular}
    };
\end{tikzpicture}

     } % 0.9
        \caption{Codes (128,64), comparison with different (128,64) codes from \cite{128_64} with 200 BP iterations, results from our previous work \cite{GNBP_Learning} and Choukroun et al. \cite{Choukroun2024b}. We observe that both learned codes from our work and the work from Choukroun et al. \cite{Choukroun2024b} exhibit a less steep slope at high SNRs compared to the AR3A code - which remains current state-of-the-art LDPC code for this size - that could indicate earlier entry into the error floor region. This characteristic could be related to the training methodology used in both approaches and suggests that optimizing for high SNR performance may require specialized training procedures or adaptive error rate policies as the model converges.}
        \label{fig:12864}
    \end{figure*}
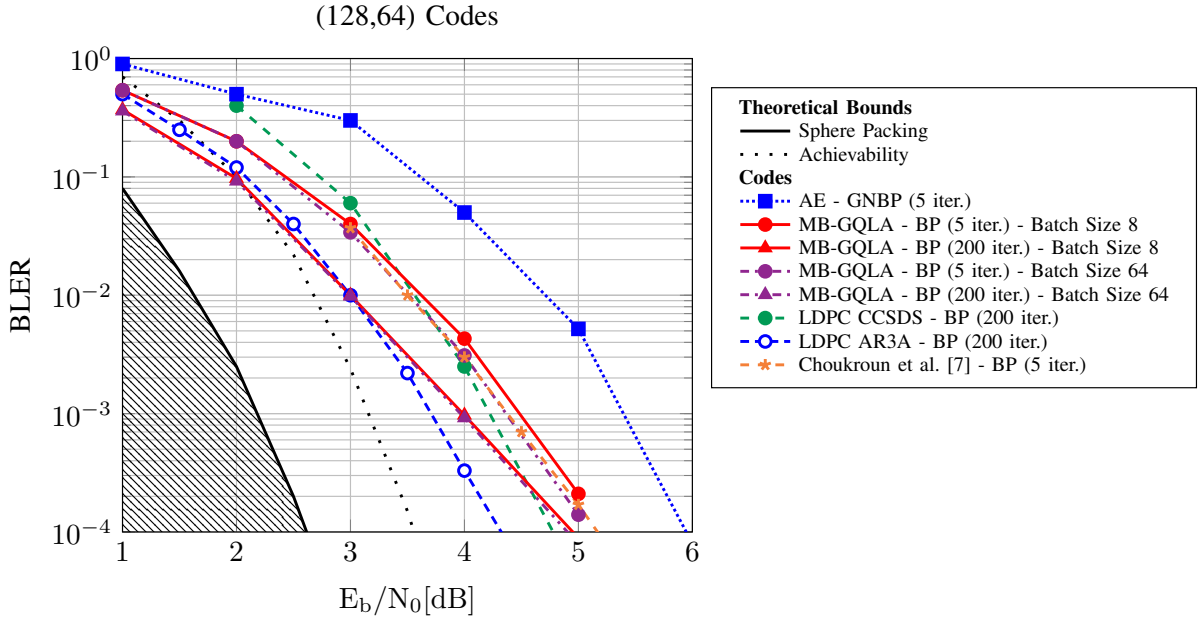%\end{figure}

    To assess the scalability of our approach, we extend our analysis to a higher code size of (128,64). We benchmark our results against the CCSDS and AR3A LDPC codes, state-of-the-art non-systematic LDPC codes for this size \cite{128_64}. Additionally, we consider the Sphere Packing (similar to the Metaconverse) and Achievability bounds from \cite{128_64} as theoretical reference points.

    A key aspect of our investigation is the impact of the number of BP iterations on performance. It is well-established that the iteration count significantly influences the efficacy of such iterative algorithms. Figure~\ref{fig:12864} illustrates these performance comparisons. Our study yields notable findings: the proposed model evaluated with only 5 BP iterations, although based on systematic codes, achieves performance comparable to the CCSDS code with 200 BP iterations. Furthermore, when our code is allowed 200 BP iterations, it exhibits performance on par with the AR3A codes under the same iteration count.

    For additional context, we include results from Choukroun et al. \cite{Choukroun2024b}, who also explore ML-based code design with 5-iteration BP decoding at this code length. Our approach achieves comparable performance, further validating the effectiveness of learning-based methods for code design.
    
    These results validate the scalability of our proposed approach and highlight its competitiveness with state-of-the-art codes, particularly in the low to moderate $\mathrm{E_b/N_0}$ regime. However, the less steep slope of our code's performance curve at higher $\mathrm{E_b/N_0}$ compared to the AR3A codes suggests room for improvement in high $\mathrm{E_b/N_0}$ scenarios. Despite this limitation, these findings underscore the promise of our learning-based approach for designing high-performance codes across various code sizes

\subsection{(64,32) Codes - Cycle, Girth and Degree Analysis}
\label{section:200runs}
    In this section, we examine the properties of the learned codes in an attempt to draw hypotheses regarding their effectiveness. The following statistics on girth and degree are provided as average distributions over 200 learned codes and 10,000 random codes. To analyze the evolution of girth and degree distributions between the start and the end of the learning process, we set the density of 1's in randomly generated codes to 25\%, same as the initialization density of the learned codes.
    
    \subsubsection{Cycles and Girth}
    
        \begin{figure*}[!ht]
            \centering
            \begin{subfigure}{0.5\textwidth}
              \centering
              \includegraphics[width=1.\linewidth]{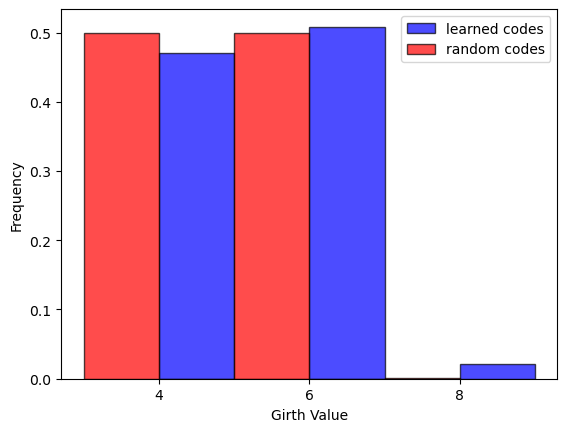}
              \caption{VN Girth}
              \label{fig:6432VNGirth}
            \end{subfigure}%
            \begin{subfigure}{0.5\textwidth}
              \centering
              \includegraphics[width=1.\linewidth]{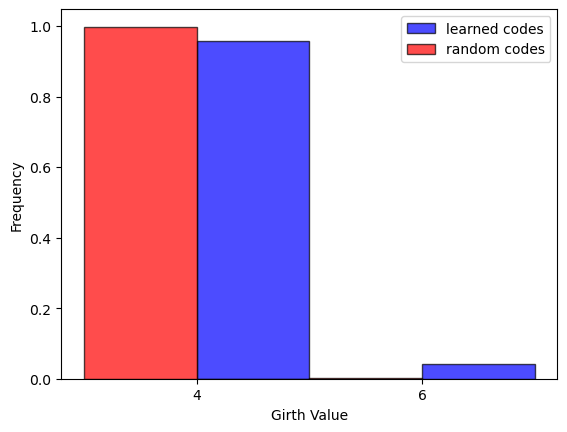}
              \caption{CN Girth}
              \label{fig:6432CNGirth}
            \end{subfigure}
            \caption{(64,32) Histogram of girth for random and learned codes.}
            \label{fig:6432Girth}
        \end{figure*}

        \begin{figure*}[htbp]
            \centering
            \begin{subfigure}{0.5\textwidth}
              \centering
              \includegraphics[width=1.\linewidth]{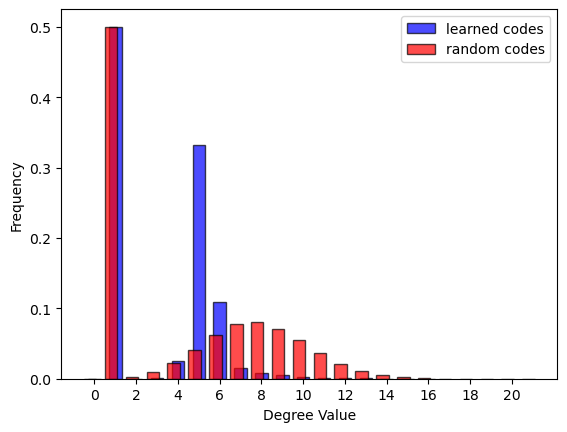}
              \caption{VN Degrees}
              \label{fig:6432VNDegrees}
            \end{subfigure}%
            \begin{subfigure}{0.5\textwidth}
              \centering
              \includegraphics[width=1.\linewidth]{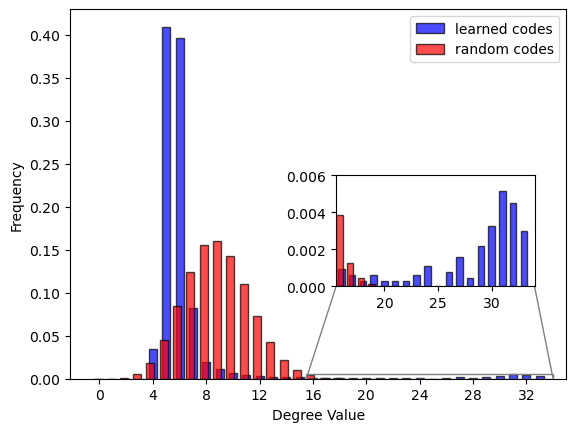}
              \caption{CN Degrees}
              \label{fig:6432CNDegrees}
            \end{subfigure}
            \caption{(64,32) Degrees distribution for random and learned codes.}
            \label{fig:6432Degrees}
        \end{figure*}
        
        A well-known limitation of BP decoders is the potential propagation of intrinsic information through cycles of the decoding graph, which can inadvertently increase confidence in erroneous LLR values. This phenomenon is particularly pronounced in dense codes with short cycles in their graph. To better understand this aspect of our learned codes, we analyze their graph properties for the (64,32) code size.
        
        We first clarify the following definitions: 
        \begin{itemize} 
            \item Code girth: The length of the shortest cycle in the complete code's graph. 
            \item Node girth: The length of the shortest cycle containing the node. 
            \item Girth histogram: The distribution of node girth values across several node of the graph. 
        \end{itemize}

        Figure~\ref{fig:6432Girth} illustrates these girth distributions for both random and learned codes. Our analysis reveals a consistent trend in the learning procedure which favors an increase in the overall girth of the code. This is evidenced by a shift in the average girth distributions for both variable nodes (VNs) and check nodes (CNs). While random code models typically display girth distributions with values up to 6 for VNs and 4 for CNs, our learned codes show a marked shift towards higher girths, with values reaching up to 8 for VNs and 6 for CNs.
        
        This increase in girth is a positive outcome, as higher girths generally correlate with improved BP decoder performance. Longer cycles in the graph can help mitigate the propagation of erroneous information, potentially leading to more reliable decoding outcomes. 
        
        % However, it's important to note that while this shift in girth distribution is noteworthy, its magnitude seems small, but we do not have reference points from state of the art to compare to.

        % \begin{figure*}[htbp]
        %     \centering
        %     \begin{subfigure}{0.5\textwidth}
        %       \centering
        %       \includegraphics[width=1.\linewidth]{figures/curves/girth/12864VNGirth.png}
        %       \caption{VN Girth}
        %       \label{fig:12864VNGirth}
        %     \end{subfigure}%
        %     \begin{subfigure}{0.5\textwidth}
        %       \centering
        %       \includegraphics[width=1.\linewidth]{figures/curves/girth/12864CNGirth.png}
        %       \caption{CN Girth}
        %       \label{fig:12864CNGirth}
        %     \end{subfigure}
        %     \caption{(128,64) Histogram of girth for random and learned codes.}
        %     \label{fig:12864Girth}
        % \end{figure*}

    \subsubsection{Node Degrees}
        Building on the girth analysis, we further examined the degree distribution of the learned codes. The degree of a node is the number of neighboring nodes directly connected to itself. The results are presented in Figure~\ref{fig:6432Degrees} for the (64,32) code size, clearly illustrating  the contrast between the degree distributions of random codes and our learned codes. Our findings reveal intriguing patterns that align with observations reported in our previous work \cite{GNBP_Learning}. On one hand, the learning procedure demonstrates a tendency to encourage overall lower density in the parity-check matrices. This preference for sparser matrices is logical, as it can help mitigate the issue of short cycles and improve the convergence of BP decoding. On the other hand, we observed the emergence of a few high-degree check nodes in the learned codes. While the exact mechanism behind this dual behavior is not yet fully understood, it presents an intriguing avenue for future research. % This pattern might represent a form of structural optimization that balances the benefits of sparse codes with the error-correction capabilities offered by high-degree nodes.
        
        % It's worth noting that this phenomenon appears less pronounced in larger code sizes, such as (128,64), compared to smaller ones like (64,32). This scaling behavior is not unexpected, as larger codes provide more degrees of freedom in their structure, potentially allowing for more subtle optimizations that don't rely as heavily on extreme degree variations.
        
        % These observations not only provide insights into the inner workings of our learning algorithm but also raise interesting questions about optimal code structures for BP decoding. The emergence of these patterns through a learning process, rather than explicit design, suggests that such structures might offer inherent advantages for error correction under BP decoding.

        % \begin{figure*}[htbp]
        %     \centering
        %     \begin{subfigure}{0.5\textwidth}
        %       \centering
        %       \includegraphics[width=1.\linewidth]{figures/curves/degrees/12864VNDegrees.png}
        %       \caption{VN Degrees}
        %       \label{fig:12864VNDegrees}
        %     \end{subfigure}%
        %     \begin{subfigure}{0.5\textwidth}
        %       \centering
        %       \includegraphics[width=1.\linewidth]{figures/curves/degrees/12864CNDegrees.png}
        %       \caption{CN Degrees}
        %       \label{fig:12864CNDegrees}
        %     \end{subfigure}
        %     \caption{(128,64) Degrees distribution for random and learned codes}
        %     \label{fig:12864Degrees}
        % \end{figure*}

\section{Conclusion}
\label{section:conclusion}
In this paper, we presented a novel approach to learning linear block codes optimized for BP decoding using a simple auto-encoder architecture. Our method hinges on a few critical hyper-parameters that significantly influence the system's ability to learn high-performing codes. 

Our learned codes have demonstrated significant performance improvements, surpassing our previous results and several state-of-the-art codes in terms of BLER. For instance, in the case of (64,32) codes, we achieved approximately 1 dB gain compared to our previous architecture. When compared to state-of-the-art codes, our proposed model, although based on systematic codes, reaches performance levels comparable to the standardized non-systematic CCSDS LDPC code of size (128,64), while requiring 40 times fewer decoding iterations.

To rigorously assess the efficacy of our learning techniques, we developed a methodology for comparison with random search approaches. This comparative analysis showed that our code learning techniques consistently outperformed random search methods in a statistically significant manner, demonstrating for the first time the value of using ML and gradient descent-based techniques to learn error-correcting codes. We also introduced gradient quantization techniques that maintain binary-valued weights throughout the training process. While these techniques offer potential benefits in terms of interpretability and update efficiency, our results suggest that the overall system design and hyper-parameter selection play a more crucial role in achieving the observed performance improvements. Additionally, we investigated potential design characteristics such as the distribution of girth and node degrees to explain the performance of these codes, confirming observations made in previous studies.

Several promising avenues for future research emerge from this work. First, the simplification of the model architecture and the relatively small number of hyper-parameters entice us to investigate further the parametrization of the learning process and corresponding results. This investigation could help identify learning policies tailored for specific code lengths, rates and channel characteristics (including dynamic hyper-parameter adjustment during training).

Secondly, we hypothesize that numerical stability in the computation graph is a critical requirement to the learning of BP-decoded codes. This underlying assumption encourages us to study the impact of the scaling factor and that of the gradient normalization techniques beyond the use of the sign function. This also prompts us to also investigate simpler message updates rules, such as the min-sum approach.

Finally, extending this work to non-systematic linear block codes remains an open challenge.
\section{Open Source Code}

The source code for these studies can be found at https://github.com/Orange-OpenSource/QNBP

\bibliographystyle{IEEEtran}
\bibliography{references}

\end{document}